%% file: main.tex
\documentclass[letterpaper,twocolumn,10pt]{article}
\usepackage{usenix2019_v3}

\usepackage{tikz}
\usepackage{amsmath}
\usepackage{color}
\usepackage{booktabs}
\usepackage{xspace}
\usepackage{comment}
\usepackage{graphicx}
\usepackage{subcaption}
\usepackage{authblk} 
\usepackage{amsmath}
\usepackage{enumitem}
\usepackage{amssymb}
\usepackage{floatrow}
\newfloatcommand{capbtabbox}{table}[][\FBwidth]
\usepackage[export]{adjustbox}
\usepackage[font=small,labelfont=bf, skip=5pt]{caption}
\newcommand{\argroup}{Rgroup\xspace}
\newcommand{\argroups}{Rgroups\xspace}
\newcommand{\adgroup}{Dgroup\xspace}

\newcommand{\heart}{HeART\xspace}
\newcommand{\pacemaker}{{\sc pacemaker}\xspace}

\newcommand{\afr}{AFR\xspace}
\newcommand{\mttdl}{MTTDL\xspace}

\newcommand{\mttr}{MTTR\xspace}

\newcommand{\expand}{RUp\xspace}
\newcommand{\contract}{RDn\xspace}

\newcommand{\contracted}{RDn transitioned\xspace}
\newcommand{\expansion}{\expand transition\xspace}
\newcommand{\contraction}{\contract transition\xspace}
\newcommand{\probname}{transition overload\xspace}

\newcommand{\clusterone}{Cluster1\xspace}

\newcommand{\initiator}{proactive-transition-initiator\xspace}
\newcommand{\Initiator}{Proactive-transition-initiator\xspace}
\newcommand{\selector}{\argroup-planner\xspace}
\newcommand{\Selector}{\argroup-planner\xspace}
\newcommand{\executor}{transition-executor\xspace}
\newcommand{\Executor}{Transition-executor\xspace}
\newcommand{\ratelimit}{rate-limit\xspace}

\newcommand{\Earlywarning}{Threshold-\afr}
\newcommand{\transitionmethodone}{Type 1\xspace}
\newcommand{\transitionmethodtwo}{Type 2\xspace}
\newcommand{\peakio}{peak-IO\xspace}
\newcommand{\peakiocap}{peak-IO-cap\xspace}
\newcommand{\avgio}{average-IO\xspace}

\newcommand{\Dadapt}{Disk-adaptive\xspace}
\newcommand{\dadapt}{disk-adaptive\xspace}
\newcommand{\Dadaptred}{\Dadapt redundancy\xspace}
\newcommand{\dadaptred}{\dadapt redundancy\xspace}
\newcommand{\tio}{transition IO\xspace}
\newcommand{\io}{IO\xspace}
\newcommand{\admin}{administrator\xspace}
\newcommand{\admins}{\admin{s}\xspace}
\newcommand{\spacesavings}{space-savings\xspace}

\newcommand{\toleratedafr}{tolerated-AFR\xspace}
\newcommand{\thresholdafr}{threshold-AFR\xspace}

\newcommand{\Toleratedafr}{Tolerated-AFR\xspace}

\newcommand{\trickledeped}{trickle-deployed\xspace}
\newcommand{\stepdeped}{step-deployed\xspace}
\newcommand{\trickledepment}{trickle-deployment\xspace}
\newcommand{\stepdepment}{step-deployment\xspace}

\newcommand{\Trickledeped}{Trickle-deployed\xspace}

\newcommand{\usefullifephase}{useful life phase\xspace}
\newcommand{\usefullifephases}{\usefullifephase{}s\xspace}

\newcommand{\enginename}{redundancy orchestrator\xspace}

\newcommand{\clusteronesize}{450K\xspace}
\newcommand{\clustertwosize}{160K\xspace}
\newcommand{\clusterthreesize}{350K\xspace}
\newcommand{\backblazesize}{110K\xspace}

\newcommand{\paramk}{$k$\xspace}
\newcommand{\paramn}{$n$\xspace}

\newcommand{\numcanaries}{$C$\xspace}

\newcommand{\datanodemanager}{DNMgr\xspace}
\newcommand{\datanode}{DN\xspace}
\newcommand{\namenode}{NN\xspace}
\newcommand{\hdfsrepo}{\url{https://github.com/thesys-lab/pacemaker-hdfs.git}}

\newcommand{\Saurabh}[1]{{\color{blue} Saurabh says: #1}}
\newcommand{\saurabh}[1]{{\color{blue} Saurabh says: #1}}
\newcommand{\greg}[1]{{\color{orange} Greg says: #1}}
\newcommand{\rashmi}[1]{{\color{red} Rashmi says: #1}}
\newcommand{\francisco}[1]{{\color{magenta} Francisco says: #1}}
\newcommand{\suhas}[1]{{\color{purple} \textbf{Suhas says}: #1}}
\newcommand{\jason}[1]{{\color{magenta} Jason says: #1}}

\newcommand{\radd}[1]{\textcolor{brown}{\textbf{Rashmi added}:\ {#1}}}
\usepackage[normalem]{ulem}
\usepackage{soul} 

\renewcommand{\Saurabh}[1]{}
\renewcommand{\saurabh}[1]{}
\renewcommand{\greg}[1]{}
\renewcommand{\rashmi}[1]{}
\renewcommand{\jason}[1]{}
\renewcommand{\francisco}[1]{}
\renewcommand{\suhas}[1]{}

\renewcommand{\radd}[1]{{{#1}}}

\newcommand{\csk}[1]{{\color{blue} C-R sk says: #1}}

\newcommand{\crv}[1]{{\color{red} C-R rv says: #1}}

\newcommand{\cgg}[1]{{\color{orange} C-R gg says: #1}}

\newcommand{\cfm}[1]{{\color{purple} C-R fm says: #1}}

\newcommand{\css}[1]{{\color{brown} C-R ss says: #1}}

\newcommand{\cjy}[1]{{\color{pink} C-R jy says: #1}}
\newcommand{\cradd}[1]{{\color{magenta} ********* C-R ADDITION: #1 *********}}
\renewcommand{\st}[1]{{{\textcolor{blue}{\sout{#1}}}}}

\renewcommand{\st}[1]{}

\renewcommand{\csk}[1]{}

\renewcommand{\crv}[1]{}

\renewcommand{\cgg}[1]{}

\renewcommand{\cfm}[1]{}

\renewcommand{\css}[1]{}

\renewcommand{\cjy}[1]{}
\renewcommand{\cradd}[1]{{#1}}

\usepackage{titlesec}
\usepackage{titling}

\titlespacing{\section}{0pt}{5pt plus 2pt minus 1pt}{2pt plus 1pt minus 2pt}
\titlespacing{\subsection}{0pt}{4pt plus 2pt minus 1pt}{2pt plus 2pt minus 2pt}
\titlespacing{\subsubsection}{0pt}{3pt plus 1pt minus 1pt}{1pt plus 2pt minus 2pt}

\microtypecontext{spacing=nonfrench}
\usepackage[available,functional,reproduced]{usenixbadges}

\begin{document}

\date{\vspace{-1.2em}}

\title{\Large \bf \textit{PACEMAKER} \\Avoiding HeART attacks in storage clusters with disk-adaptive redundancy\vspace{-1em}}

\author{
    Saurabh Kadekodi, Francisco Maturana, Suhas Jayaram Subramanya, Juncheng Yang, \vspace{-10pt}  \\
    K. V. Rashmi, Gregory R. Ganger\\
    \textit{Carnegie Mellon University}
}

\maketitle 
\pagenumbering{gobble}
\pagestyle{empty}
\vspace*{-12mm}

\begin{abstract} 
\input{sections/abstract}
\end{abstract}
\input{sections/introduction_v4.tex} 
\input{sections/new_background.tex}
\input{sections/datasets.tex}
\input{sections/design_v3.tex}
\input{sections/hdfs.tex}
\input{sections/evaluation_v2.tex}
\input{sections/related_work.tex}
\input{sections/conclusion.tex}
\input{sections/ack.tex}
\appendix 
\input{sections/new_appendix}
\clearpage
\bibliographystyle{plain}
\bibliography{refs.bib}

\end{document}

%% file: sections/abstract.tex
Data redundancy 
provides resilience
in large-scale storage clusters,
but imposes significant cost overhead. 
Substantial \spacesavings 
can be realized
by tuning redundancy schemes to observed disk failure rates. 
However, prior design proposals for such tuning are unusable in real-world clusters, because the IO load of transitions between schemes
overwhelms the storage infrastructure (termed {\it transition overload}).

This paper analyzes 
traces for millions of disks from production systems at Google, NetApp, and Backblaze 
to expose and understand transition overload as a roadblock to \dadaptred:
\tio under existing approaches can consume 100\% cluster IO continuously for several weeks. 
Building on the insights drawn, 
we present \pacemaker, a low-overhead \dadapt \enginename. \pacemaker mitigates transition overload by (1) proactively organizing data layouts to make future transitions efficient, and (2) initiating transitions proactively in a manner that avoids urgency while not compromising on \spacesavings. 
Evaluation of \pacemaker with traces from four large (110K--450K disks) production clusters show that the transition IO requirement decreases to never needing more than 5\% cluster IO bandwidth (0.2--0.4\% on average). 
\suhas{"reduces to never needing more than" --> " can be limited to "}
\pacemaker achieves this while providing overall \spacesavings of 14--20\% and never leaving data under-protected. 
We also describe and experiment with an integration of \pacemaker into HDFS.

%% file: sections/introduction_v4.tex
\section{Introduction}
\label{sec:intro}

Distributed storage systems use data redundancy to protect data 
in the face of disk failures\cite{ford2010availability,ghemawat2003google,shvachko2010hadoop}. While it provides resilience, redundancy imposes significant cost overhead. 
Most large-scale systems today erasure code most of the data stored, instead of replicating, which helps to reduce the space overhead well below 100\%~\cite{weatherspoon2002erasure,zhang2010does,huang2012erasure,ford2010availability,rashmi2015hitchhiker,sathiamoorthy2013xoring}. 
Despite this, space overhead remains a key concern in large-scale systems since it directly translates to an increase in the number of disks and the associated increase in capital, operating and energy costs~\cite{huang2012erasure,ford2010availability,rashmi2015hitchhiker,sathiamoorthy2013xoring}. 

\begin{figure}[t!]
    \centering
    \vspace*{-4mm}
    \begin{subfigure}[t]{\textwidth}
        \centering
        \includegraphics[width=\textwidth]{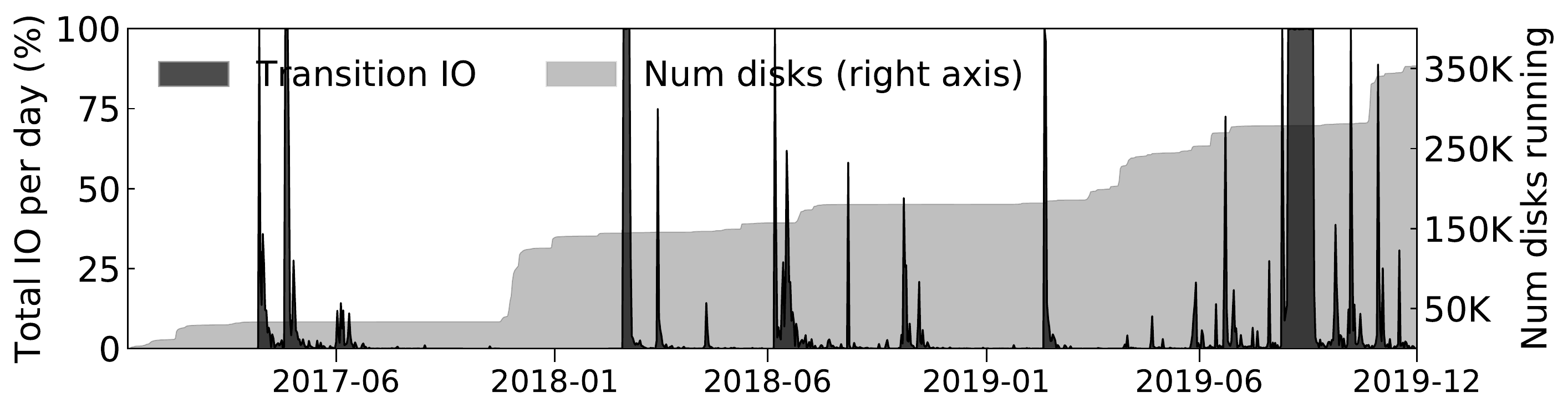}\vspace{-5pt}
        \caption{Transition IO for \heart~\cite{kadekodi2019cluster} on Google \clusterone.}
        
        \label{fig:heart_front}
    \end{subfigure}
    \begin{subfigure}[t]{\textwidth}
        \centering
        \includegraphics[width=\textwidth]{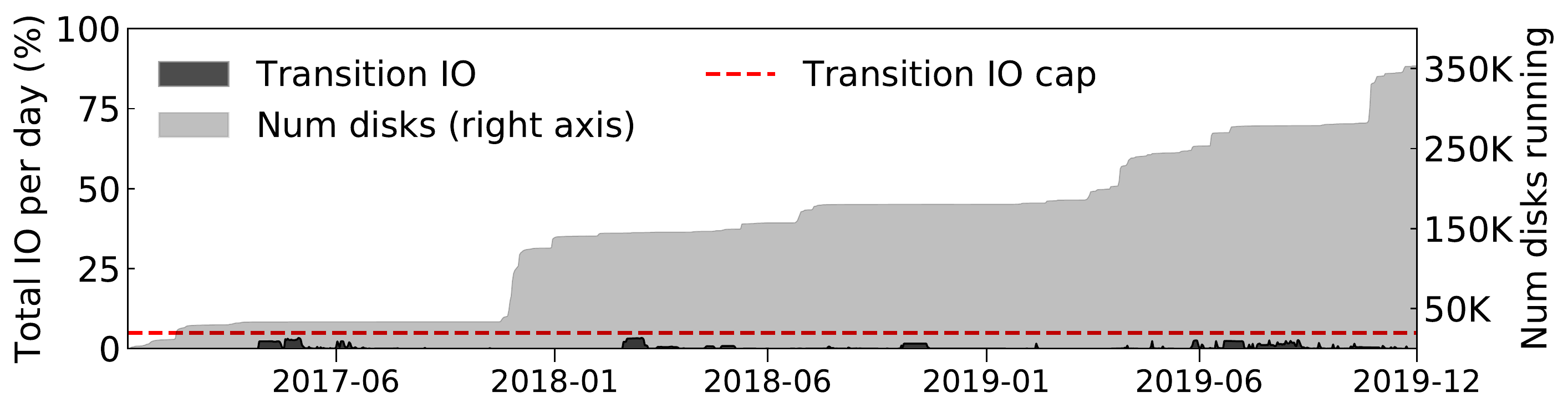}\vspace{-5pt}
        \caption{Transition IO for \pacemaker on Google \clusterone.} 
        \label{fig:pacer_front}
    \end{subfigure}
    \vspace{-.7em}
    \caption{
    Fraction of total cluster IO bandwidth needed to use disk-adaptive redundancy for a Google storage cluster's first three years.
    The state-of-the-art proposal~\cite{kadekodi2019cluster} shown in \subref{fig:heart_front} would require up to 100\% of the cluster bandwidth for extended periods, whereas \pacemaker shown in \subref{fig:pacer_front} always fits its IO under a cap (5\%).
    The light gray region shows the disk count (right Y-axis) over time. 
    }
    \label{fig:heart_vs_pacer_front}
\end{figure}

Storage clusters are made up of disks from a mix of makes/models acquired over time, and
different makes/models have highly varying failure rates\cite{ma2015raidshield,pinheiro2007failure,kadekodi2019cluster}. Despite that,
storage clusters employ a ``one-size-fits-all-disks'' approach to choosing redundancy levels, without considering failure rate differences among disks. 
Hence, space  
overhead is often inflated by overly conservative redundancy levels, chosen to ensure sufficient protection for the most failure-prone disks in the cluster.
\cradd{
Although tempting, the overhead cannot be removed by using very ``wide'' codes (which can provide high reliability with low storage overhead) for all data, due to the prohibitive reconstruction cost induced by the most failure-prone disks (more details in \S~\ref{sec:background}).
} 
An exciting alternative is to dynamically adapt redundancy choices to 
observed failure rates (\afr{s})\footnote{\afr describes the expected
fraction of disks that experience failure
in a typical year.} 
for different disks, which recent proposals suggest could substantially reduce the space overhead~\cite{kadekodi2019cluster}. 

Adapting redundancy involves dynamic transitioning of redundancy schemes, because \afr{s} must be learned from observation of deployed disks and because \afr{s} change over time due to disk aging. 
Changing already encoded data from one redundancy scheme to another, for example from an erasure code with parameters $k_1$-of-$n_1$ to $k_2$-of-$n_2$ (where $k$-of-$n$ denotes $k$ data chunks and $n-k$ parity chunks; more details in \S~\ref{sec:background}),
can be exorbitantly IO intensive.
Existing designs for \dadaptred 
are rendered unusable by overwhelming bursts of urgent transition IO when applied to real-world storage clusters.
Indeed, as illustrated in Fig.~\ref{fig:heart_front}, our analyses of production traces show extended periods 
of needing 100\% of the cluster's IO bandwidth for transitions.
We refer to this as the \textit{\probname} problem.
At its core, transition overload occurs whenever an observed \afr increase for a subset of disks requires too much urgent \tio in order to keep data safe. Existing designs for \dadaptred perform redundancy transitions as a reaction to \afr changes.
Since prior designs are reactive,
for an increase in \afr, the data is already under-protected by the time the transition to increase redundancy is issued. And it will continue to be under-protected until that transition completes. 
For example, around 2019-09 in Fig.~\ref{fig:heart_front}, data was under-protected for over a month, even though the entire cluster's IO bandwidth was used solely for redundancy transitions. Simple \ratelimit{ing} to reduce urgent bursts of IO would only exacerbate this problem causing data-reliability goals to be violated for even longer.

To understand the causes of transition overload and inform solutions, we analyse multi-year deployment and failure logs for over 5.3~million disks from Google, NetApp and Backblaze.
Two common transition overload patterns are observed.
First, sometimes disks are added in tens or hundreds over time, which we call \textit{trickle} deployments.
A statistically confident \afr observation requires thousands of disks. Thus, by the time it is known that \afr for a specific make/model and age is too high for the redundancy used, the oldest thousands of that make/model will be past that age.
At that point, all of those disks need immediate transition.
Second, sometimes disks are added in batches of many thousands, which we call \textit{step} deployments. Steps have sufficient disks for statistically confident \afr estimation. However, when a step reaches an age where the \afr is too high for the redundancy used,
\textit{all} disks of the step need immediate transition.

This paper introduces \textit{\pacemaker}, a new disk-adaptive redundancy orchestration system that exploits insights from the aforementioned analyses to eliminate the \probname problem.
\pacemaker proactively organizes data layouts 
to enable efficient transitions for each deployment pattern, reducing total transition IO by over 90\%.  
Indeed, by virtue of its reduced total \tio, \pacemaker can afford to use extra transitions to reap increased \spacesavings.
\pacemaker also
proactively initiates anticipated transitions sufficiently in advance 
that
the resulting \tio can be \ratelimit{ed} without placing data at risk.
Fig.~\ref{fig:pacer_front} provides a peek into the final result:
\pacemaker achieves \dadaptred with substantially less total \tio 
and never exceeds 
a specified transition IO cap (5\% in the graph).

We evaluate \pacemaker{}
using logs containing  all disk deployment, failure, and decommissioning events  from four production storage clusters: three \clustertwosize--\clusteronesize-disk Google clusters
and a $\approx$110K-disk cluster used for the Backblaze Internet backup service~\cite{backblaze}.
On all four clusters, \pacemaker provides \dadaptred while using less than 0.4\% of cluster IO bandwidth for transitions on average, and never exceeding the specified rate limit (e.g., 5\%) on
IO bandwidth. 
Yet, despite its proactive approach, \pacemaker loses less than 3\% of the \spacesavings as compared to to an idealized system with perfectly-timed 
and 
instant transitions.
Specifically, \pacemaker provides 14--20\% average \spacesavings 
compared to a one-size-fits-all-disks approach, without ever failing to meet the target data reliability and with no transition overload. 
We note that this is substantial savings for large-scale systems, where even a single-digit space-savings is worth the engineering effort.
For example, in aggregate, the four clusters would need $\approx$200K fewer disks. 

We also implement \pacemaker{} in HDFS, demonstrating that \pacemaker{}'s mechanisms fit into an existing cluster storage system with minimal changes.
Complementing our longitudinal evaluation using traces from large scale clusters, 
we report measurements of redundancy transitions in \pacemaker{}-enhanced HDFS via small-scale cluster experiments. \cradd{Prototype of HDFS with Pacemaker is open-sourced and is available at \hdfsrepo}.

This paper makes five primary contributions.
First, \cradd{it demonstrates that transition overload is a roadblock}
that
precludes use of 
previous disk-adaptive redundancy proposals.
Second, it presents insights into the sources of transition overload 
from longitudinal analyses of deployment and failure logs for 5.3 million disks from three large organizations.
Third, it describes \pacemaker{'s} novel techniques, designed 
based on insights drawn from these analyses, for safe disk-adaptive redundancy without transition overload.
Fourth, it evaluates \pacemaker{}'s policies for four large real-world storage clusters, demonstrating their effectiveness for a range of deployment and disk failure patterns.
Fifth, it describes 
integration of and experiments with \pacemaker{}'s techniques in HDFS, demonstrating their feasibility, 
functionality, and ease of integration into a cluster storage implementation.

%% file: sections/new_background.tex
\section{Whither \dadaptred}
\label{sec:background}

\textbf{Cluster storage systems and data reliability.}
Modern storage clusters scale to huge capacities by combining up to hundreds of thousands of storage devices into a single storage system~\cite{ghemawat2003google,shvachko2010hadoop,weil2006ceph}. 
In general, there is a metadata service that tracks data locations (and other metadata) and a large number of storage servers that each have up to tens of disks. Data is partitioned into chunks that are spread among \suhas{among -> across} the storage servers/devices.
Although hot/warm data is now often stored on Flash SSDs, cost considerations lead to the majority of data continuing to be stored on mechanical disks (HDDs) for the foreseeable future~\cite{brewer-keynote, brewer2016disks, Reinsel2018}.
For the rest of the paper, any reference to a ``device'' or ``disk'' implies HDDs. 

Disk failures are common 
and storage clusters use data redundancy to protect against irrecoverable data loss in the face of disk failures
~\cite{ghemawat2003google, huang2012erasure, pinheiro2007failure, backblaze, rashmi2013solution, rashmi2015hitchhiker,sathiamoorthy2013xoring}.
For hot data, 
often replication is used for 
performance benefits.
But, for most bulk and colder data, cost considerations have led to the use of erasure coding schemes. Under a \paramk-of-\paramn coding scheme, each set of \paramk data chunks are coupled with \paramn-\paramk ``parity chunks'' to form a ``stripe''. A \paramk-of-\paramn scheme provides tolerance to \suhas{'provides tolerance to' -> can tolerate' ?} $(n-k)$ failures with a space overhead of $\frac{n}{k}$. 
Thus, erasure coding achieves substantially lower space overhead 
for tolerating a given number of failures.
Schemes like 6-of-9 and 10-of-14 are commonly used in real-world deployments~\cite{ford2010availability, rashmi2013solution, rashmi2015hitchhiker, sathiamoorthy2013xoring}. Under erasure coding, additional work is involved in recovering from a device failure. To reconstruct a lost chunk, 
\paramk remaining chunks from the stripe must be read. 

\textbf{The redundancy scheme selection problem.}
The reliability of data stored redundantly is often quantified as \textit{mean-time-to-data-loss} (\mttdl)~\cite{gibson1992redundant}, which essentially captures the average time until more than the tolerated number of chunks are lost. 
\mttdl is calculated using the disks' \afr 
and its \textit{mean-time-to-repair} (\mttr). 

Large clusters are built over time, and hence usually consist of a mix of disks belonging to multiple makes/models depending on which options were most cost effective at each time.
\afr{} values vary significantly between makes/models and disks of different ages~\cite{kadekodi2019cluster, ma2015raidshield, pinheiro2007failure, schroeder2007disk}. Since disks have different \afr{s}, computing \mttdl of a candidate redundancy scheme for a large-scale storage cluster is often difficult. 

The \mttdl{} equations can still be used to guide decisions, as long as a sufficiently high \afr{} value is used. For example, if the highest \afr{} value possible for any deployed make/model at any age  
is used, the computed \mttdl{} will be a lower bound. So long as the lower bound on \mttdl meets the target \mttdl, the data is adequately reliable. 
Unfortunately, the range of possible \afr{} values in a large storage cluster is generally quite large (over an order of magnitude)~\cite{kadekodi2019cluster, ma2015raidshield, pinheiro2007failure, schroeder2016flash}. Since the overall average is closer to the lower end of the \afr range, the highest \afr value is a conservative over-estimate for most disks. The resulting \mttdl{}s are thus loose lower bounds, prompting decision-makers to use a one-size-fits-all scheme with excessive redundancy leading to wasted space.

\cradd{Using wide schemes with large number of parities (e.g., 30-of-36) can achieve the desired \mttdl while keeping the storage overhead low enough to make \dadaptred appear not worth the effort. 
But, while this might seem like a panacea, 
wide schemes in high-AFR regimes cause significant increase in failure reconstruction IO traffic.
The failure reconstruction IO is derived by multiplying the \afr with the number of data chunks in each stripe. Thus, if either of these quantities are excessively high, or both are moderately high, it can lead to overwhelmingly high failure reconstruction IO. In addition, wide schemes also result in higher tail latencies for individual disk reconstructions because of having to read from many more disks. Combined, these reasons prevent use of wide schemes for all data all the time from being a viable solution for most systems.}

\textbf{Disk-adaptive redundancy.} 
Since the problem arises from using 
a single \afr{} value, a promising alternative is to adapt  
redundancy for subsets of disks with similar \afr{s}.
A recent proposal,  
heterogeneity-aware redundancy tuner (\heart)~\cite{kadekodi2019cluster}, suggests treating subsets of deployed disks with different \afr{} characteristics differently.
Specifically, \heart adapts redundancy of each disk by observing its failure rate on the fly\footnote{Although it may be tempting to use \afr{} values taken from manufacturer's specifications, several studies have shown that failure rates observed in practice often do not match those~\cite{pinheiro2007failure, schroeder2007disk, schroeder2016flash}.} depending on its make/model and its current age.
It is well known that \afr of disks follow a ``bathtub'' shape with three distinct phases of life: 
\afr is high in ``infancy'' (1-3 months), low and stable during its ``useful life'' (3-5 years), and high during the ``wearout'' (a few months before decommissioning). 
\heart uses a default (one-size-fits-all) redundancy scheme for each new disk's infancy. It then dynamically changes the redundancy to a scheme adapted to the observed useful life \afr for that disk's make/model, and then dynamically changes back to the default scheme at the end of useful life.
The per-make/model useful life redundancy schemes typically have much lower space overhead than the default 
scheme. This suggests 
the ability to maintain target \mttdl with many fewer disks (i.e., lower cost).

Although exciting, the design of \heart overlooks a crucial element:  the IO cost associated with changing the redundancy schemes. Changing already encoded data under one erasure code to another can be exorbitantly IO intensive. Indeed, 
our evaluation of \heart on real-world storage cluster logs reveal extended periods where data safety is at risk and where 100\% cluster IO bandwidth is consumed for scheme changes. We call this problem \textit{\probname}. 

\cradd{An enticing solution that might appear to mitigate \probname is to adapt redundancy schemes only by removing parities in low-\afr regimes and adding parities in high-\afr regimes. 
While this solution eliminates transition \io when reducing the level of redundancy, it does only marginally better when redundancy needs to be increased, because new parity creation cannot avoid reading all data chunks from each stripe. 
What makes this worse is that transitions that increase redundancy are time-critical, since delaying them would miss the \mttdl target and leave the data under-protected. Moreover, addition / removal of a parity chunk massively changes the stripe's \mttdl compared to addition / removal of a data chunk. 
For example, a 6-of-9 \mttdl is 10000$\times$ higher than 6-of-8 \mttdl, but is only 1.5$\times$ higher than 7-of-10 \mttdl.
\afr changes would almost never be large enough to safely remove a parity, given default schemes like 6-of-9, eliminating almost all potential benefits of disk-adaptive redundancy.
}

This paper analyzes disk deployment and failure data from large-scale production clusters to discover sources of \textit{transition overload} and informs the design of a solution.
It then describes and evaluates \pacemaker, which realizes the dream of safe disk-adaptive redundancy without transition overload.

%% file: sections/datasets.tex
\section{Longitudinal production trace analyses}
\label{sec:traceanalysis}
\label{sec:production}

\begin{figure*}[t]
    \centering
    \begin{subfigure}[t]{0.38\textwidth}
        \centering
        \includegraphics[width=\textwidth]{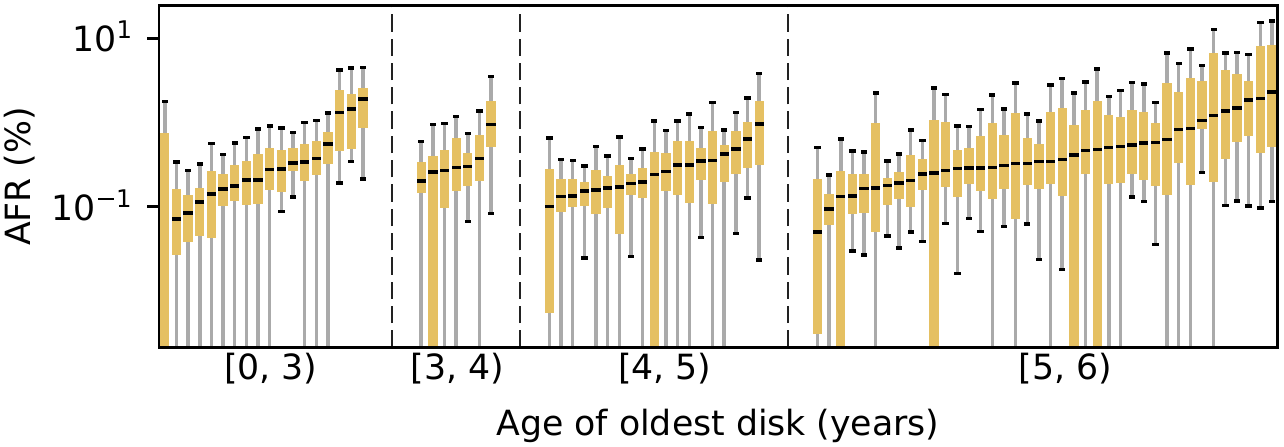}\vspace{-5pt}
        \caption{Spread of make/model \afr{}s.}
        \label{fig:netapp_afr_variance}
    \end{subfigure}
    \begin{subfigure}[t]{0.27\textwidth}
        \centering
        \includegraphics[width=\textwidth]{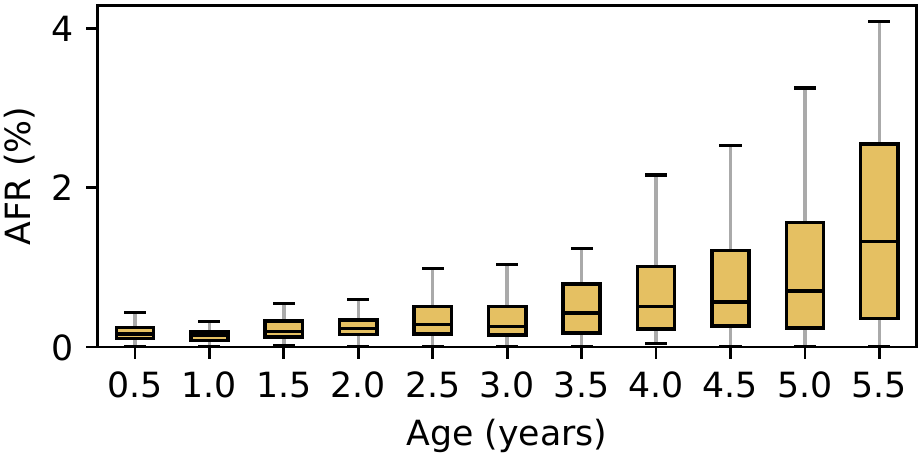}\vspace{-5pt}
        \caption{\afr distribution over disk life.}
        \label{fig:netapp_gradual_wearout}
    \end{subfigure}
    \begin{subfigure}[t]{0.33\textwidth}
        \centering
        \includegraphics[width=\textwidth]{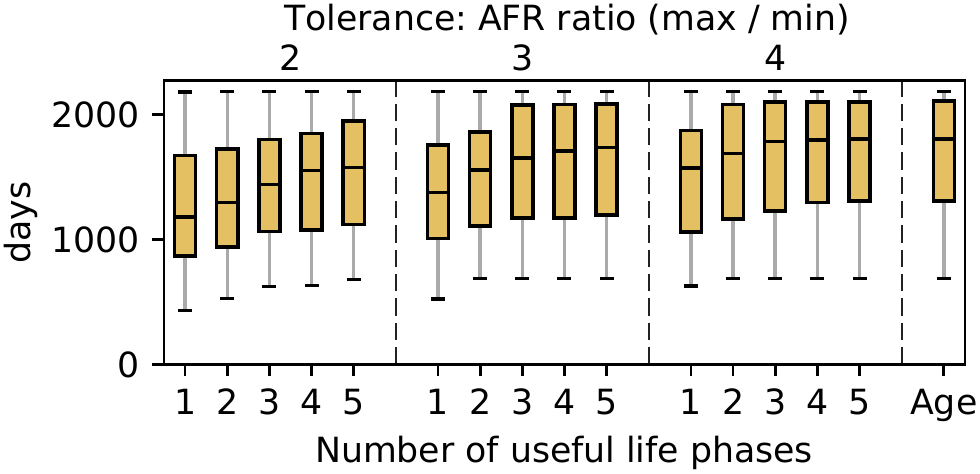}\vspace{-5pt}
        \caption{Approximate useful-life length.}
        \label{fig:netapp_multiple_useful_lives}
    \end{subfigure}
    \vspace{-5pt}
    \caption{\ref{fig:netapp_afr_variance}
        \afr spread for over 50 makes/models from NetApp binned by the age of the oldest disk. Each box corresponds to a unique make/model, and at least 10000 disks of each make/model were observed (outlier \afr{} values omitted).
        \ref{fig:netapp_gradual_wearout}
        Distribution of \afr{} calculated over consecutive non-overlapping six-month periods for NetApp disks, showing the gradual rise of \afr{} with age (outliers omitted).
        \ref{fig:netapp_multiple_useful_lives}
        Approximation of useful life length for NetApp disks for 1-5 consecutive phases of useful life and three different tolerance levels.
        \vspace{-15pt}.}
    \label{fig:dataset_fig}
\end{figure*}

This section presents an analysis of multi-year disk reliability logs and deployment characteristics of 5.3 million HDDs, covering over 60 makes/models from real-world environments. Key insights presented here shed light on the sources of transition overload and challenges / opportunities for a robust \dadaptred solution.

\textbf{The data.} 
Our largest dataset comes from NetApp and contains information about disks deployed in filers (file servers).
Each filer reports the health of each disk periodically (typically once a fortnight) 
using their AutoSupport~\cite{lancaster2001measuring} system. 
We analyzed the data for a subset of their deployed disks, which included over 50 makes/models and over 4.3 million disks total. 
As observed in previous studies~\cite{kadekodi2019cluster,pinheiro2007failure,schroeder2007disk}, we observe well over an order of magnitude difference between the highest and lowest useful-life \afr{}s 
(see Fig.~\ref{fig:netapp_afr_variance}).

Our other datasets come from large storage clusters deployed at Google and the Backblaze Internet backup service.
Although the basic disk characteristics (e.g., \afr heterogeneity and its behavior discussed below) are similar to the NetApp dataset, these datasets also capture the evolution and behavior in our target context (large-scale storage clusters), and thus are also used in the evaluation detailed in (\S\ref{sec:evaluation}).
The particular Google clusters were selected based on their longitudinal data availability, but were not otherwise screened for favorability.

For each cluster, the multi-year log  records (daily) all disk deployment, failure, and decommissioning events from birth of the cluster until the date of the log snapshot.
Google Cluster1's disk population over
three years included $\approx$\clusterthreesize disks of 7 makes/models.
Google Cluster2's population over 2.5~years included $\approx$\clusteronesize disks of 4 makes/models. 
Google Cluster3's population over 3 years included $\approx$\clustertwosize disks of 3 makes/models.
The Backblaze cluster's population since 2013 included $\approx$\backblazesize disks of 7 makes/models. 

\subsection{Causes of transition overload}
\label{sec:production.deployment}

{\bf Disk deployment patterns.}
We observe disk deployments occurring in two distinct patterns, which
we label \textit{\textbf{trickle}} and \textit{\textbf{step}}.
\Trickledeped disks are added to a cluster frequently (weekly or even daily) over time by the tens and hundreds.
For example, the slow rise in disk count seen between 2018-01 and 2018-07 in Fig.~\ref{fig:heart_vs_pacer_front} represents a series of \trickledepment{s}.
In contrast, a \stepdepment introduces many thousands of disks into the cluster ``at once'' (over a span of a few days), followed by potentially months of no new \stepdepment{s}.
The sharp rises in disk count around 2017-12 and 2019-11 in Fig.~\ref{fig:heart_vs_pacer_front} represent \stepdepment{s}.

A given cluster may be entirely \trickledeped (like the Backblaze cluster), entirely \stepdeped (like Google Cluster2), or a mix of the two (like Google Cluster1 and Cluster3). Disks of a step are typically of the same make/model. 

{\bf Learning \afr curves online.}
\Dadaptred involves learning the \afr curve for each make/model by observing failures among deployed disks of that make/model. Because \afr{} is a statistical measure, the larger the population of disks observed at a given age, the lower is the uncertainty in the calculated \afr{} at that age. We have found that a few thousand disks need to be observed to obtain sufficiently accurate \afr{} measurements.

{\bf Transition overload for \trickledeped disks.}
Since \trickledeped disks are deployed in tiny batches over time, several months can pass before the required number of disks of a new make/model are past any given age.
Thus, by the time the required number of disks can be observed at the age that is eventually identified as having too-high an \afr and requiring increased redundancy, data on the older disks will have been left under-protected for months. 
And, the thousands of already-older disks need to be immediately transitioned to a stronger redundancy scheme, together with the newest disks to reach that age. This results in 
\probname. 

{\bf Transition overload for \stepdeped disks.}
Assuming that they are of the same make/model, a batch of \stepdeped disks will have the same age and \afr, and indeed represent a large enough population for confident learning of the \afr curve as they age. 
But, this means that all of those disks will reach \afr values together, as they age.
So, when their \afr rises to the point where the redundancy must be increased to keep data safe, all of the disks must transition together to the new safer redundancy scheme.
Worse, if they are the first disks of the given make/model deployed in the cluster, which is often true in the clusters studied, then the system adapting the redundancy will learn of the need only when the age in question is reached.
At that point, all data stored on the entire batch of disks is unsafe and needs immediate transitioning. This results in \probname. 

\subsection{Informing a solution}
\label{sec:production.afr}
Analyzing the disk logs has exposed a number of observations that provide hope and guide the design of \pacemaker.
The \afr curves we observed deviate substantially from the canonical representation where infancy and wearout periods are identically looking and have high \afr values, and \afr in useful life is flat and low throughout.

\textbf{\afr{s} rise gradually over time with no clear wearout.}
\afr curves generally exhibit neither a flat useful life phase nor a sudden transition to so-called wearout.
Rather, in general, it was observed that \afr curves rise gradually as a function of disk age.
Fig.~\ref{fig:netapp_gradual_wearout} shows the gradual rise in \afr over six month periods of disk lifetimes. Each box represents the \afr of disks whose age corresponds to the six-month period denoted along the X-axis.
\afr curves for individual makes/models (e.g., Figs.~\ref{fig:main_afr_1} and~\ref{fig:main_afr_2}) are consistent with this aggregate illustration.
Importantly, none of the over 60 makes/models from Google, Backblaze and NetApp displayed sudden onset of wearout.

Gradual increases in \afr, rather than sudden onset of wearout, suggests that one could anticipate a \stepdeped batch of disks approaching an \afr threshold.
This is one foundation on which \pacemaker's proactive transitioning approach rests.

\textbf{Useful life could have multiple phases.} 
Given the gradual rise of \afr{s}, 
useful life can be decomposed into multiple, piece-wise constant phases.
\cradd{
Fig.~\ref{fig:netapp_multiple_useful_lives} shows an approximation of the length of useful life when multiple phases are considered.
Each box in the figure represents the distribution over different make/models of the approximate length of useful life. Useful life is approximated by considering the longest period of time which can be decomposed into multiple consecutive phases (number of phases indicated by the bottom X-axis) such that the ratio between the maximum and minimum AFR in each phase is under a given tolerance level (indicated by the top X-axis). The last box indicates the distribution over make/models of the age of the oldest disk, which is an upper bound to the length of useful life. As shown by Fig.~\ref{fig:netapp_multiple_useful_lives}, the length of useful life can be significantly extended (for all tolerance levels) by considering more than one phase. Furthermore, the data show that a small number of phases suffice in practice, as the approximate length of useful life changes by little when considering four or more phases.
}

\textbf{Infancy often short-lived.}
Disks may go through (potentially) multiple rounds of so-called ``burn-in'' testing. The first tests may happen at the manufacturer's site. There may be additional burn-in tests done at the deployment site allowing most of the infant mortality to be captured before the disk is deployed in production.
For the NetApp  and Google disks, we see the \afr drop sharply and plateau by 20 days for most of the makes/models. 
In contrast, the Backblaze disks display a slightly longer and higher \afr during infancy, which can be directly attributed to their less aggressive on-site burn-in. 

\pacemaker{'s} design is heavily influenced from these learnings, as will be explained in the next section.

%% file: sections/design_v3.tex
\section{Design goals}
\label{sec:design.goals}

\pacemaker is an IO efficient \enginename for storage clusters that support \dadaptred. 
Before going into the design goals for \pacemaker, we first chronicle a disk's lifecycle, introducing the terminology that will be used in the rest of the paper (defined in Table~\ref{tab:terminology}). 

\textbf{Disk lifecycle under \pacemaker.} 
Throughout its life, each disk under \pacemaker simultaneously belongs to a \textit{\adgroup} and an \textit{\argroup}.
There are as many \adgroup{s} in a cluster as there are unique disk makes/models. \argroup{s} on the other hand are a function of redundancy schemes and placement restrictions. 
Each \argroup has an associated redundancy scheme, and its data (encoded stripes) must reside completely within that \argroup{'s} disks. 
Multiple \argroup{s} can use the same redundancy scheme,
but no stripe 
may span across \argroup{s}.
The \adgroup of a disk never changes, but a disk may transition through multiple \argroup{s} during its lifetime. 
At the time of deployment (or ``birth''), 
the disk belongs to \textit{\argroup{0}}, and is termed as an \textit{unspecialized disk}. Disks in \argroup{0} use the default redundancy scheme, i.e. the conservative one-scheme-fits-all scheme used in storage clusters that do not have \dadaptred.
{The redundancy scheme employed for a disk (and hence its \argroup) changes via \textit{transitions}.} 
The first transition any disk undergoes is an \textit{\contraction}. A \contraction changes the disk's
\argroup to one with lower redundancy, i.e.\ more optimized for space. 
Whenever the disk departs from \argroup{0}, it is termed as a \textit{specialized disk}. Disks depart from \argroup{0} at the end of their infancy. Since infancy is short-lived (\S\ref{sec:production.afr}), \pacemaker only considers one \contraction for each disk.

\begin{table}[tp]
    \footnotesize
    \begin{tabular}{*2l}
    \toprule
    \emph{Term} & \emph{Definition} \\
    \midrule
    \textbf{\adgroup} & Group of disks of the same make/model.\\ 
    \textbf{Transition} & The act of changing the redundancy scheme.\\
    \textbf{\contraction} & Transition to a lower level of redundancy.\\ 
    \textbf{\expansion} & Transition to a higher level of redundancy.\\ 
    \textbf{\peakiocap} & IO bandwidth cap for transitions.\\
    \textbf{\argroup} & Group of disks using the same redundancy \\
    & with placement restricted to the group of disks.\\
    \textbf{\argroup{0}} & \argroup using the default one-scheme-fits-all\\
    & redundancy used in storage clusters today.\\
    \textbf{Unspecialized disks} & Disks that are a part of \argroup{0}.\\
    \textbf{Specialized disks} & Disks that are not part of \argroup{0}.\\
    \textbf{Canary disks} & First few thousand disks of a \trickledeped\\
    & \adgroup used to learn \afr curve.\\
    \textbf{\Toleratedafr} & Max \afr for which redundancy scheme meets\\
    & reliability constraint.\\
    \textbf{\Earlywarning} & The \afr threshold crossing which triggers \\
    & an \expansion for \stepdeped disks.\\
    \bottomrule
    \hline
    \end{tabular}\vspace{-5pt}
    \caption{Definitions of \pacemaker{'s} terms. 
    \vspace{-8pt}} 
    \label{tab:terminology}
\end{table}

The first \contraction occurs at the start of the disk's useful life, and marks the start of its specialization period. As explained in \S\ref{sec:production.afr}, a disk may experience multiple useful life phases. \pacemaker performs a transition at the start of each \usefullifephase. After the first (and only) \contraction, each subsequent transition is an \textit{\expansion}. An \expansion changes the disk's \argroup to one with higher redundancy, i.e. less optimized for space, but the disk is still considered a specialized disk unless the \argroup that the disk is being \expansion{ed} to is \argroup{0}. The \spacesavings (and thus cost-savings) associated with \dadaptred are proportional to the fraction of life the disks remain specialized for.

\textbf{Key decisions.} 
To adapt redundancy throughout a disk's lifecycle as chronicled above, 
three key decisions related to transitions must be made 
\vspace{-5pt}
\begin{enumerate}[noitemsep]
    \item {\textbf{\textit{When should the disks transition?}}}
    \item {\textbf{\textit{Which \argroup should the disks transition to?}}} 
    \item {\textbf{\textit{How should the disks transition?}}}
\end{enumerate}\vspace{-5pt}

\textbf{Constraints.} The above decisions need to be taken such that a set of constraints are met. 
An obvious constraint,
central to any storage system, is that of data reliability. The \textit{reliability constraint} mandates that all data must always meet a predefined target \mttdl. 
Another important constraint is the \textit{failure reconstruction IO constraint}. This constraint bounds the IO spent on data reconstruction of failed disks, which as explained in \S\ref{sec:background} is proportional to \afr and scheme width. This is why wide schemes cannot be used for all disks all the time, but they can be used for low-\afr regimes of disk lifetimes \radd{(as discussed in \S\ref{sec:background})}.

Existing approaches to \dadaptred make their decisions on the basis of only these constraints~\cite{kadekodi2019cluster}, but fail to consider the equally important \textit{IO caused by redundancy transitions}. Ignoring this causes the \probname problem, which proves to be a show-stopper for \dadaptred systems.
\pacemaker treats \tio as a first class citizen by taking it into account for each of its three key decisions. As such, \pacemaker enforces carefully designed constraints on \tio as well.

\textbf{Designing \io constraints on transitions.}
Apart from serving foreground IO requests, a storage cluster performs numerous background tasks like scrubbing and load balancing~\cite{bairavasundaram2007analysis, schroeder2010understanding, oprea2010clean}. 
Redundancy management is also a background task. In current storage clusters, 
redundancy management tasks predominantly consist of performing data redundancy (e.g.\ replicating or encoding data) 
and reconstructing data of failed or otherwise unavailable disks.
Disk-adaptive redundancy systems add redundancy
transitions to the list of \io-intensive background tasks.

There are two goals for background tasks:
Goal 1: they are not too much work, and Goal 2: they interfere as little as possible with foreground \io. 
\pacemaker 
applies two IO constraints on background transition tasks to achieve these goals: 
(1) \textit{\avgio constraint} and (2) \textit{\peakio constraint}.
The \avgio constraint achieves Goal 1
by allowing storage \admins to specify a cap on the fraction of the \io bandwidth of a disk that can be used for transitions over its lifetime. 
For example, if a disk can transition in 1 day using 100\% of its \io bandwidth, then an \avgio constraint of 1\% would mean that the disk will transition at most once every 100 days.
The \peakio constraint achieves Goal 2
by allowing storage \admins to specify the 
peak rate (defined as the \textit{\peakiocap}) at which transitions can occur so as to limit their interference with foreground traffic.
Continuing the previous example, if the \peakiocap is set at 5\%, the disk that would have taken 1~day to transition at 100\% \io bandwidth would now take at least 20~days. The \avgio constraint and the \peakiocap can be configured based on how busy the cluster is. For example, a cluster designed for data archival would have a lower foreground traffic, compared to a cluster designed for serving ads or recommendations. Thus, low-traffic clusters can set a higher \peakiocap resulting in faster transitions and potentially increased \spacesavings.

\textbf{Design goals.} 
The key design goals are to {answer the three questions related to transitions such that the \spacesavings are maximized and the following constraints are met}: (1) reliability constraint on all data all the time, (2) failure reconstruction \io constraint on all disks all the time, (3) \peakio constraint on all disks all the time, and (4) \avgio constraint on all disks over time.

\begin{figure}[t]
    \centering
    \includegraphics[width=0.9\textwidth]{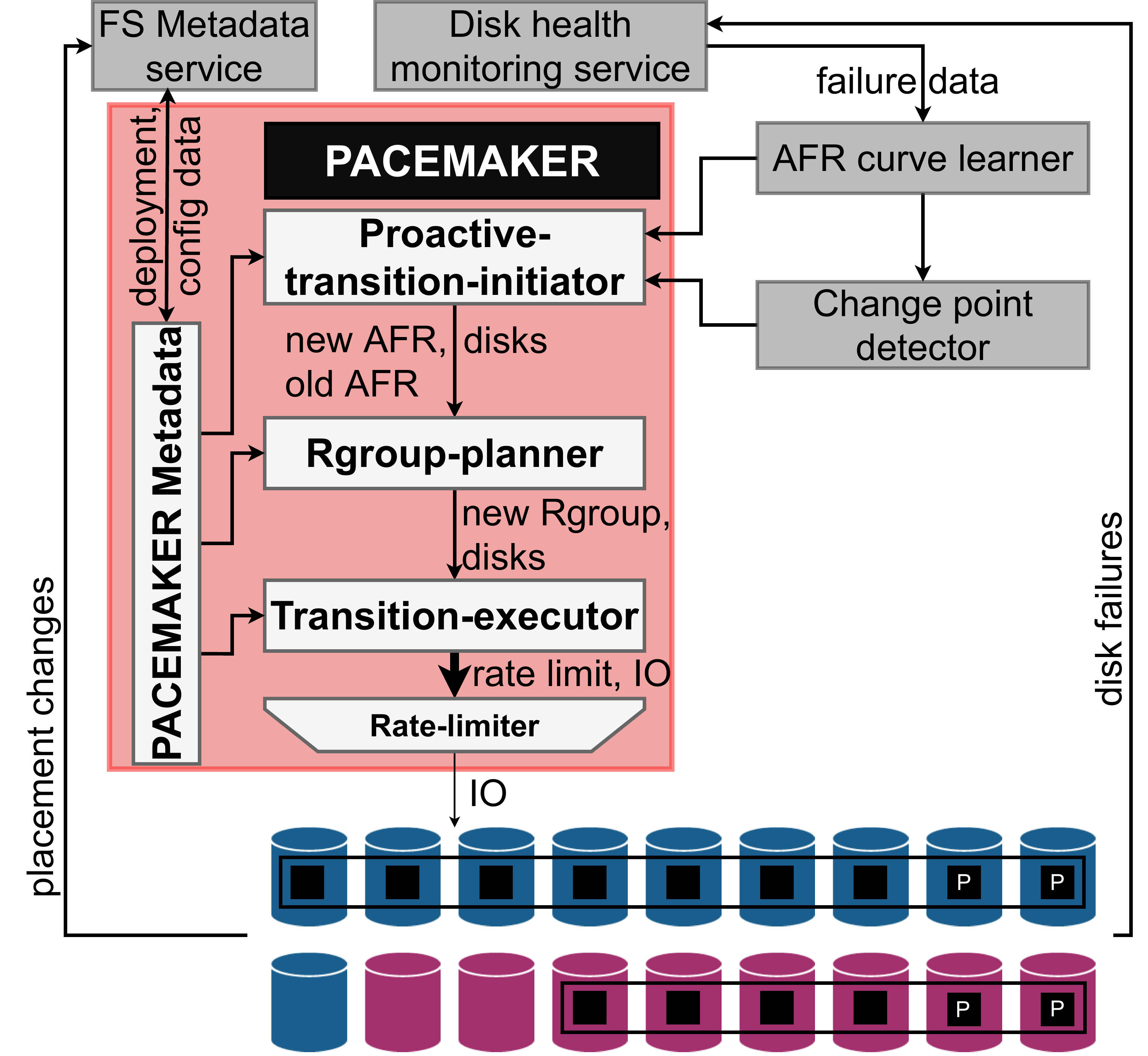}\vspace{-5pt}
    \caption{\pacemaker architecture.
    \vspace{-7pt}}
    \label{fig:architecture}
\end{figure}

\section{Design of \pacemaker}
Fig.~\ref{fig:architecture} shows the high level architecture of \pacemaker and how it interacts with some other components of a storage cluster.
The three main components of \pacemaker correspond to the three key decisions that the system makes as discussed in \S\ref{sec:design.goals}. 
The first main component of \pacemaker is the \textit{\textbf{\initiator}}
(\S\ref{subsec:initiator}), which determines when to transition disks using the \afr curves and the disk deployment information. 
The information of the transitioning disks 
and their observed \afr is passed to the {\textbf{\textit{\selector}}} (\S\ref{subsec:planner}), which chooses the \argroup to which the disks should transition.
The \selector passes the information of the transitioning disks and the target \argroup to the \textit{\textbf{\executor}} (\S\ref{sec:design.executor}). The \executor addresses how to transition the disks to the planned \argroup in the most \io-efficient way. 

Additionally, \pacemaker also maintains its own \textbf{\textit{metadata}} and a simple \textbf{\textit{\ratelimit{er}}}. \pacemaker metadata interacts with all of \pacemaker{'s} components and also the storage cluster's metadata service. It maintains various configuration settings of a \pacemaker installation along with the disk deployment information that guides transition decisions.  The \ratelimit{er} \ratelimit{s} the \io load generated by any transition as per administrator specified limits.
Other cluster components external-to-\pacemaker that inform it are the \textit{\afr curve learner} and the \textit{change point detector}. As is evident from their names, these components learn the \afr curve\footnote{The \afr estimation methodology employed
is detailed in appendix~\ref{app:afr_estimation}. 
}
of each \adgroup and identify change points for redundancy transitions.
The \afr curve learner receives failure data 
from the \textit{disk health monitoring service}, which monitors the disk fleet and maintains their vitals.

\subsection{\Initiator}\label{subsec:initiator}
\Initiator{'s} role is to determine \textit{when to transition the disks}.
Below we explain \pacemaker{'s} methodology for making this decision for the two types of transitions (\contract and \expand) and the two types of deployments (step and trickle). 

\subsubsection{
\textbf{Deciding when to \contract a disk}
}
Recall that a disk's first transition is an \contraction. As soon as \initiator observes (in a statistically accurate manner) that the \afr has decreased sufficiently, and is stable, it performs an \contraction from the default scheme (i.e., from \argroup{0}) employed in infancy to a more space-efficient scheme. This is the only \contraction in a disk's lifetime.

\subsubsection{\textbf{Deciding when to \expand a disk}} \label{subsec:when-to-rup}
\cradd{\expansion{s} are performed either when there are too few disks in any \argroup such that data placement is heavily restricted (which we term \textit{purging an \argroup}), or when there is a rise in \afr such that the reliability constraint is (going to be) violated. Purging an \argroup involves \expansion{ing} all of its disks to an \argroup with higher redundancy. This transition isn't an imminent threat to reliability, and therefore can be done in a relaxed manner without violating the reliability constraint as explained in \S\ref{subsec:rate-limiting}.}

However, most \expansion{s} in a storage cluster are done in response to a rise in \afr. These are challenging with respect to meeting IO constraints due to the associated risk of violating the reliability constraints whenever the \afr rises beyond the \afr tolerated by the redundancy scheme (termed \textit{\toleratedafr}). 

In order to be able to safely \ratelimit the IO load due to \expansion{s}, \pacemaker takes a \textit{proactive} approach. The key is in determining when to initiate a {proactive} \expansion such that the transition can be completed before the \afr crosses the \toleratedafr, while adhering to the IO and the reliability constraints without compromising much on \spacesavings. To do so, the \initiator assumes that its transitions will proceed as per the peak-IO constraint, which is ensured by the \executor.
\pacemaker{'s} methodology for determining when to initiate a {proactive} \expansion 
is tailored differently for trickle versus for step deployments, since they raise different challenges.

\textbf{Trickle deployments.}\label{subsec:canaries}
For \trickledeped disks, \pacemaker considers two category of disks: (1) first disks to be deployed from any particular \trickledeped \adgroup, and (2) disks from that \adgroup that are deployed later.

\pacemaker labels the first \numcanaries deployed disks of a \adgroup as \textit{canary} disks, where \numcanaries is a configurable, high enough number of disks to yield statistically significant \afr{} observations. For example, based on our disk analyses,
we observe that \numcanaries in low thousands (e.g., 3000) is sufficient.
The canary disks of any \adgroup are the first to undergo the various phases of life for that \adgroup, and these observations are used to learn the \afr curve for that \adgroup. 
The \afr value for the \adgroup at any particular age is not known (with statistical confidence) until all canary disks go past that age. 
Furthermore, due to the trickle nature of the deployment, the canary disks would themselves have been deployed over weeks if not months. Thus, \afr for the canary disks can be ascertained only in retrospect. 
\pacemaker never changes the redundancy of the canary disks to avoid them from ever violating the reliability constraint. This does not significantly reduce \spacesavings, since \numcanaries is expected to be small relative to the total number of disks of a \adgroup (usually in the tens of thousands).

The disks that are deployed later in any particular \adgroup are easier to handle, since the \adgroup{'s} \afr curve would have been learned by observing the canaries. 
Thus, the date at which a disk among the later-deployed disks needs to \expand 
to meet the reliability constraints is known in advance by the \initiator, which it uses to issue proactive \expansion{s}. 

\textbf{Step deployments.}\label{subsec:early_warning}
Recall that in a step deployment, most disks of a \adgroup may be deployed within a few days. So, canaries are not a good solution, as they would provide little-to-no advance warning about how the \afr curve's rises would affect most disks. 

\pacemaker{'s} approach to handling \stepdepment{s} is based on two properties: (1) Step-deployments have a large number of disks deployed together, leading to a statistically accurate \afr estimation;
(2) \afr curves based on a large set of disks tend to exhibit gradual, rather than sudden, \afr increases as the disk ages (\S\ref{sec:production.afr}).
\pacemaker leverages these two properties to employ a simple \textit{early warning} methodology to predict a forthcoming need to 
\expansion a step well in advance. 
Specifically, \pacemaker sets a threshold, termed \textit{\thresholdafr},
which is a (configurable) fraction of the \toleratedafr of the current redundancy scheme employed. 
For \stepdepment{s}, when the observed \afr crosses the \thresholdafr, the \initiator initiates  
a proactive \expansion. 

\subsection{\Selector}
\label{sec:design.what}
\label{subsec:planner}
The \selector{'s} role is to determine \textit{which \argroup should disks transition to}. 
This involves making two \textit{interdependent} choices: (1) the redundancy scheme to transition into, (2) whether or not to create a new \argroup.

\textbf{Choice of the redundancy scheme.}
At a high level, the \selector first uses a set of selection criteria to arrive at a set of viable schemes. It further narrows down the choices by filtering out the schemes that are not worth transitioning to when the \tio and \io constraints are accounted for.

\textit{Selection criteria for viable schemes.} 
Each viable redundancy scheme has to satisfy the following criteria in addition to the reliability constraint: each scheme (1) must satisfy the minimum number of simultaneous failures per stripe (i.e., $n-k$); (2) must not exceed the maximum allowed stripe dimension ($k$); (3) must have its expected failure reconstruction \io 
(\afr $\times$ $k$ $\times$ disk-capacity) be no higher than was assumed possible for \argroup{0} (since disks in \argroup{0} are expected to have the highest \afr);
(4) must have a recovery time in case of failure (\mttr) that does not exceed the maximum \mttr (set by the \admin when selecting the default redundancy scheme for \argroup{0}).

\textit{Determining if a scheme is worth transitioning to.}
Whether the \io cost of transitioning to a scheme is worth it or not and what \spacesavings can be achieved by that transition 
is a function of the number of days disks will remain in that scheme (also known as \textit{disk-days}). This, in turn, depends on 
(1) when the disks enter the new scheme, and (2) how soon disks will require another transition out of that scheme. 

The time it takes for the disks to enter the new scheme is determined by the transition IO and the \ratelimit. 
When the disks will transition out of the target \argroup is dependent on the future and can only be estimated. 
For this estimation,
the \selector needs to estimate the number of days the \afr curve will remain below the threshold that forces a transition out. This needs different strategies for the two deployment patterns (trickle and step). 

Recall that \pacemaker knows the \afr curve for \trickledeped disks (from the canaries) in advance. 
Recall 
that \stepdeped disks have the property that the \afr curve learned from them is statistically robust and tends to exhibit gradual, as opposed to sudden \afr increases. 
The \selector leverages these properties to estimate the future \afr behavior based on the recent past. Specifically, it takes the slope of the \afr curve in the recent past\footnote{
\pacemaker uses a 60 day (configurable) sliding window with an Epanechnikov kernel, which gives more weight to \afr changes in the recent past~\cite{hastie2009kernel}.%
} and uses that to project the \afr curve rise in the future. 

The number of disk-days 
in a scheme for it to be worth transitioning to is dictated by the IO constraints. For example, let us consider a disk running under \pacemaker that requires a transition, and \pacemaker is configured with an \avgio constraint of 1\% and a \peakiocap of 5\%. Suppose the disk requires 1 day to complete its transition at 100\% IO bandwidth. With the current settings, \pacemaker will only consider an \argroup worthy of transitioning to (assuming it is allowed to use all 5\% of its \io bandwidth) if at least 80 disk-days are spent after the disk entirely transitions to it (since transitioning to it would take up to 20 days at the allowed 5\% IO bandwidth).

From among the viable schemes that are worth transitioning to based on the IO constraints, the \selector chooses the one that provides the highest \spacesavings. 

\textbf{Decision on \argroup creation.}
\cradd{\argroup{s} cannot be created arbitrarily. This is because every \argroup adds \textit{placement restrictions}, since all chunks of a stripe have to be stored on disks belonging to the same \argroup. Therefore, \selector creates a new \argroup only when (1) the resulting placement pool created by the new \argroup is large enough to overcome traditional placement restrictions such as ``no two chunks on the same rack\footnote{\cradd{Inter-cluster fault tolerance remains orthogonal to and unaffected by \pacemaker.}}'', 
and (2) the \spacesavings achievable by the chosen redundancy scheme is sufficiently greater than using an existing (less-space-efficient) \argroup.

The disk deployment pattern also affects \argroup formation. While the rules for whether to form an \argroup remain the same for trickle and \stepdeped disks, mixing disks deployed differently impacts the transitioning techniques that can be used for eventually transitioning disks out of that \argroup. This in turn affects how the \io constraints are enforced. Specifically, for trickle deployments, creating an \argroup for each set of transitioning disks would lead to too many small-sized \argroup{s}. So, for \trickledepment{s}, the \selector creates a new \argroup for a redundancy scheme if and only if one does not exist already. Creating \argroup{s} this way 
will also ensure that enough disks (thousands) will go into it to satisfy placement restrictions. Mixing disks from different \trickledepment{s} in the same \argroup does not impact the \io constraints, because \pacemaker optimizes the transition mechanism for few disks transitioning at a time, as is explained in \S\ref{sec:design.executor}. For \stepdepment{s}, due to the large fraction of disks that undergo transition together, having disks from multiple steps, or mixing \trickledeped disks within the same \argroup, creates adverse interactions (discussed in \S\ref{sec:design.executor}). Hence, the \selector creates a new \argroup for each \stepdepment, even if there already exists one or more \argroup{s} that employ the chosen scheme.
Each such \argroup will contain many thousands of disks to overcome traditional placement restrictions. Per-step \argroup{s} also extend to the \argroup with default redundancy schemes, implying a per-step \argroup{0}. Despite having clusters with disk populations as high as 450K disks, \pacemaker{'s} restrained \argroup creation led to no cluster ever having more than 10 \argroup{s}.}

\cradd{\textbf{Rules for purging an \argroup.}
An \argroup may be purged for having too few disks. This can happen when too many of its constituent disks transition to other \argroup{s}, or they fail, or they are decommissioned leading to difficulty in fulfilling placement restrictions. If the \argroup to be purged is made up of \trickledeped disks, the \selector will choose to \expansion disks to an existing \argroup with higher redundancy while meeting the \io constraints. For \stepdepment{s}, purging implies \expansion{ing} disks into the more-failure-tolerant RGroup (RGroup0) that may include \trickledeped disks.}

\subsection{\Executor}\label{sec:design.executor}
The \executor{'s} role is to determine \textit{how to transition the disks}. This involves choosing (1) the most \io-efficient technique to execute that transition, and
(2) how to \ratelimit the transition at hand. 
Once the transition technique is chosen, the \executor executes the transition via the \ratelimit{er} as shown in Fig.~\ref{fig:architecture}.

\textbf{Selecting the transition technique.}
Suppose the data needs to be conventionally re-encoded from a $k_{cur}$-of-$n_{cur}$ scheme to a $k_{new}$-of-$n_{new}$ scheme. The IO cost of conventional re-encoding involves reading--re-encoding--writing all the stripes whose chunks reside on each transitioning disk. This amounts to a read IO of ${k_{cur}}\times$disk-capacity (assuming almost-full disks), and a write IO of ${k_{cur}}\times$disk-capacity$\times\frac{n_{new}}{k_{new}}$ for a total IO $>2\times{k_{cur}}\times$disk-capacity for each disk.

In addition to conventional re-encoding, \pacemaker supports two new approaches to changing the redundancy scheme for disks and selects the most efficient option for any given transition. 
The best option depends on the fraction of the \argroup being transitioned at once.

\textit{\transitionmethodone (Transition by emptying disks)}.
\label{sec:ideas.transcoding_via_decom}
If a small percentage of an \argroup{'s} disks are being transitioned, it is more efficient to retain the contents of the transitioning disks in that \argroup rather than re-encoding. Under this technique, the data stored on transitioning disks are simply moved (copied) to other disks within the current \argroup. This involves reading and writing (elsewhere) the contents of the transitioning disks. Thus, the IO of transitioning via \transitionmethodone is at most 2$\times$disk-capacity, independent of scheme parameters, and therefore at least ${k_{cur}}\times$ cheaper than conventional re-encoding. 

\transitionmethodone can be employed whenever there is sufficient free space available to move the contents of the transitioning disks into other disks in the current \argroup. 
Once the transitioning disks are empty, they can be removed from the current \argroup and added to the new \argroup as ``new'' (empty) disks. 

\textit{\transitionmethodtwo (Bulk transition by recalculating parities).}
If a large fraction of disks in an \argroup need to transition together, it is more efficient to transition the entire \argroup rather than only the disks that need a transition at that time. 
Most cluster storage systems use systematic codes\footnote{In systematic codes, the data chunks are stored in unencoded form. This helps to avoid having to decode for normal (i.e., non-degraded-mode) reads.}~\cite{hdfs-ec, calder2011windows, ford2010availability, muralidhar2014f4}, wherein transitioning an entire \argroup involves only calculating and storing new parities and deleting the old parities. Specifically, the data chunks have to be only read for computing the new parities, but they do not have to be re-written. In contrast, if only a part of the disks are transitioned, some fraction of the data chunks also need to be re-written. Thus, the IO cost for transitioning via \transitionmethodtwo 
involves a read IO of $\frac{k_{cur}}{n_{cur}}\times$disk-capacity, and a write IO of only the new parities, which amounts to a total IO of $\frac{n_{new}-k_{new}}{k_{new}}\times\frac{k_{cur}}{n_{cur}}\times$disk-capacity for each disk in the \argroup. This is at most $2\times\frac{k_{cur}}{n_{cur}}\times$disk-capacity, which makes it at least $n_{cur}\times$ cheaper than conventional re-encoding.

\textit{Selecting the most efficient approach for a transition.}
For any given transition, the \executor selects the most IO-efficient of all the viable approaches. 
Almost always, trickle-deployed disks use \transitionmethodone{} because they transition a-few-at-a-time, and step-deployed disks use \transitionmethodtwo{} because \selector maintains each step in a separate \argroup.

\textbf{Choosing how to rate limit a transition.}\label{subsec:rate-limiting}
Irrespective of the transitioning techniques, the \executor has to resolve the competing concerns of maximizing \spacesavings and minimizing risk of data loss via fast transitions, and minimizing foreground work interference by slowing down transitions so as to not overwhelm the foreground IO. Arbitrarily slowing down a transition to minimize interference is only possible when the transition is not in response to a rise in \afr. 
This is because a rising \afr hints at the data being under-protected if not transitioned to a higher redundancy soon. In \pacemaker, a transition without an \afr rise occurs either when disks are being \contraction{ed} at the end of infancy, or when they are being \expansion{ed} because the \argroup they belong to is being purged. For all the other \expansion{s}, \pacemaker carefully chooses how to rate limit the transition.

Determining how much bandwidth to allow for a given 
transition could be difficult, given that other transitions may be in-progress already or may be initiated at any time (we do observe concurrent transitions in our evaluations).
So, to ensure that the aggregate IO of all ongoing transitions conforms to the \peakiocap cluster-wide, \pacemaker limits each transition to the \peakiocap  within its \argroup.
For trickle-deployed disks, which share \argroup{s}, the rate of transition initiations is consistently a small percentage of the shared \argroup, allowing disk emptying to proceed at well below the \peakiocap. 
For step-deployed disks, this is easy for \pacemaker, since a step only makes one transition at a time and its IO is fully contained in its separate \argroup.
The \executor's approach to managing \peakio on a per-\argroup{} basis is also why the \initiator can safely assume a \ratelimit of the \peakiocap without consulting the \executor. 
\cradd{If there is a sudden \afr increase that puts data at risk, \pacemaker is designed to ignore its IO constraints to continue meeting the reliability constraint---this safety valve was never needed for any cluster evaluated.}

After finalizing the transitioning technique, the \executor performs the necessary IO for transitioning disks (read, writes, parity recalculation, etc.). We find that the components required for the \executor are already present and adequately modular in existing distributed storage systems. In \S\ref{sec:hdfs}, we show how we implement \pacemaker in HDFS with minimal effort.

\cradd{Note that this design is for the common case where storage clusters are designed for a single dedicated storage service. Multiple distinct distributed storage services independently using the same underlying devices would need to coordinate their use of bandwidth (for their non-transition related load as well) in some way, which is outside the scope of this paper.}

%% file: sections/hdfs.tex
\section{Implementation of \pacemaker in HDFS}
\label{sec:hdfs}

We have implemented 
a prototype
of \pacemaker for the Hadoop distributed file system (HDFS)~\cite{shvachko2010hadoop}. HDFS is a popular 
open source distributed file system, widely employed in the industry for storing large volumes of data. 
We use HDFS v3.2.0,
which natively supports erasure coding. \cradd{Prototype of HDFS with Pacemaker is open-sourced and is available at \hdfsrepo.}

\textbf{Background on HDFS architecture.}
HDFS has a central metadata server called Namenode (\namenode, akin to the master node) and a collection of servers containing the data stored in the file system, called Datanodes (\datanode, akin to worker nodes). 
Clients interact with the \namenode only to perform operations on file metadata (containing a collection of the \datanode{s} that store the file data).
Clients directly request the data from the \datanode{s}.
Each \datanode 
stores data on its local drives using a local file system.

\begin{figure}[t]
\centering
\includegraphics[width=\textwidth]{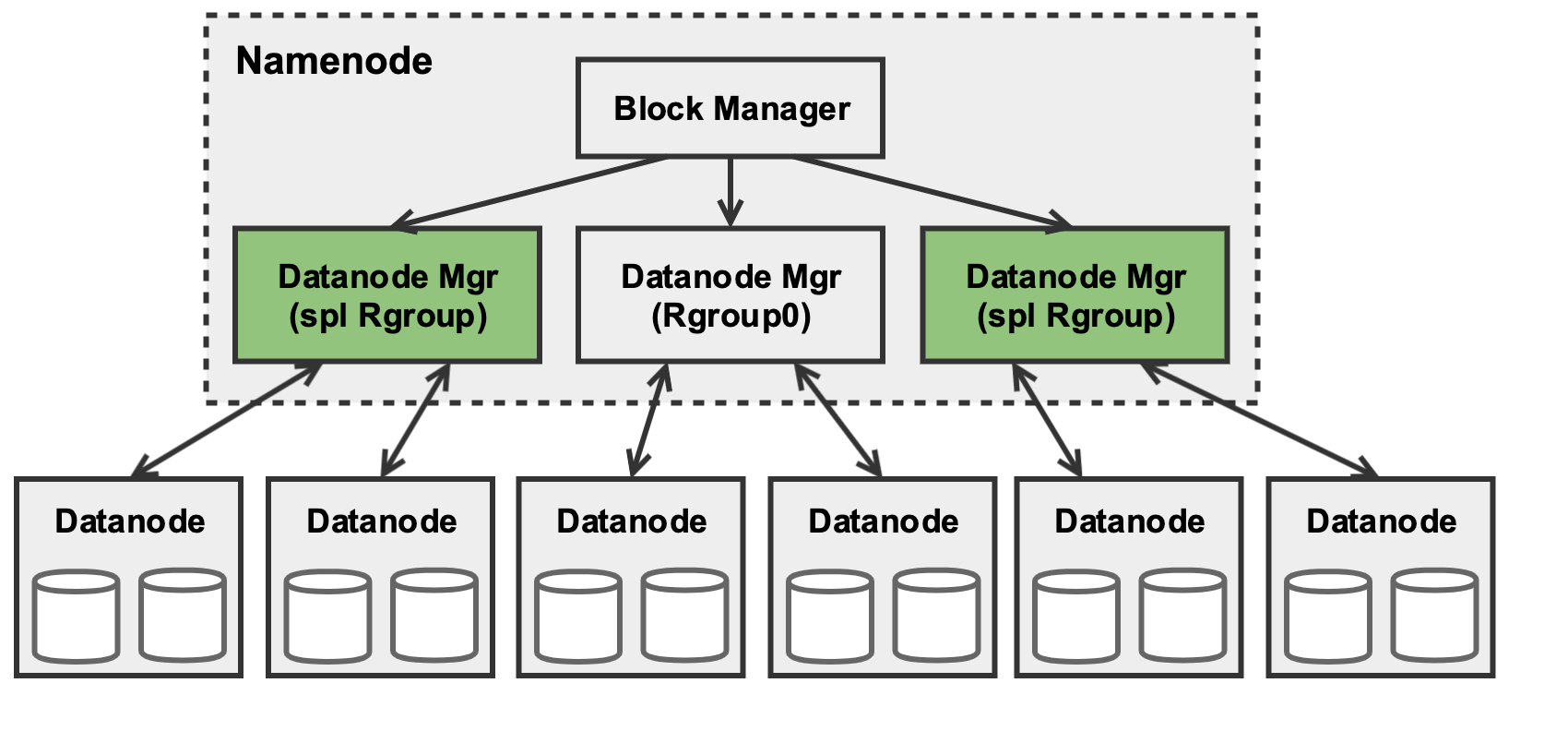}\vspace{-3pt}
\caption{\pacemaker{-enhanced} HDFS architecture.%
\vspace{-5pt}}
\vspace{-1.1em}
\label{fig:preact_hdfs_architecture}
\end{figure}

\textbf{Realizing \argroups in HDFS.}
This design makes a simplifying assumption that all disks belonging to a \datanode are of the same \adgroup and are deployed together (this could be relaxed easily).
Under this simplifying assumption, conceptually, an \argroup would consist of a set of \datanode{s} that need to be managed independent of other such sets of \datanode{s} as shown in Fig~\ref{fig:preact_hdfs_architecture}. 

The \namenode maintains a DatanodeManager (\datanodemanager), which is a gateway for the \namenode to interact
with the \datanode{s}. The \datanodemanager maintains a list of the \datanode{s}, along with their usage statistics. The \datanodemanager also contains a HeartBeatManager (HrtBtMgr) which handles the periodic keepalive heartbeats from \datanode{s}. A natural mechanism to realize \argroup{s} in HDFS is to have one \datanodemanager per \argroup. 
Note that the sets of \datanode{s} belonging to the different \datanodemanager{s} are mutually exclusive. 
Implementing \argroup{s} with multiple \datanodemanager{s} has several advantages.

\textit{Right level of control and view of the system.} Since the \datanodemanager resides below the block layer, when the data needs to be moved for redundancy adaptations, the logical view of the file remains unaffected. Only the mapping from HDFS blocks to \datanode{s} gets updated in the inode. 
The statistics maintained by the \datanodemanager can be used to balance load across \argroup{s}.

\textit{Minimizing changes to the HDFS architecture and maximizing re-purposing of existing HDFS mechanisms.} This design obviates the need to change HDFS's block placement policy, since it is implemented at the \datanodemanager level. Block placement policies 
are notoriously hard to get right. Moreover, block placement decisions are affected by fault domains and network topologies, both of which are orthogonal to \pacemaker's goals, and thus best left untouched. 
Likewise, the code for reconstruction of data from a failed \datanode need not be touched, since all of the reads (to reconstruct each lost chunk) and writes (to store it somewhere else) will occur within the set of nodes managed by its \datanodemanager. Existing mechanisms for adding / decommissioning nodes managed by the \datanodemanager can be re-purposed to implement \pacemaker{'s} \transitionmethodone transitions (described below).

\textit{Cost of maintaining multiple \datanodemanager{s} is small.} Each \datanodemanager maintains two threads: a HrtBtMgr and a DNAdminMgr. The former tracks and handles heartbeats from each \datanode, and the latter monitors the \datanode{s} for performing decommissioning and maintenance. 
The number of \datanodemanager threads in the \namenode will increase from two to 2$\times$ the number of \argroup{s}. Fortunately, even for large clusters, we observe that 
the number of \argroup{s} would not exceed the low tens (\S\ref{sec:hdfs_eval}). The \namenode is usually a high-end server compared to the \datanode{s}, and an additional tens of threads shouldn't affect performance.

\textbf{\argroup transitions in HDFS.}
\label{subsec:transitioning_datanodes}
An important part of \pacemaker functionality is transitioning \datanode{s} between \argroups. Recall from \S\ref{sec:ideas.transcoding_via_decom} that one of \pacemaker{'s} preferred way of transitioning disks across \argroup{s} is by emptying the disks. In HDFS, the planned removal of a \datanode from a HDFS cluster is called decommissioning. 
\pacemaker re-uses decommissioning to remove a \datanode from the set of \datanode{s} managed by one \datanodemanager and then adds it to the set managed by another, effectively transitioning a \datanode from one \argroup to another.

\begin{figure*}[t]
    \begin{subfigure}[t]{0.77\textwidth}
        \includegraphics[width=\textwidth,left]{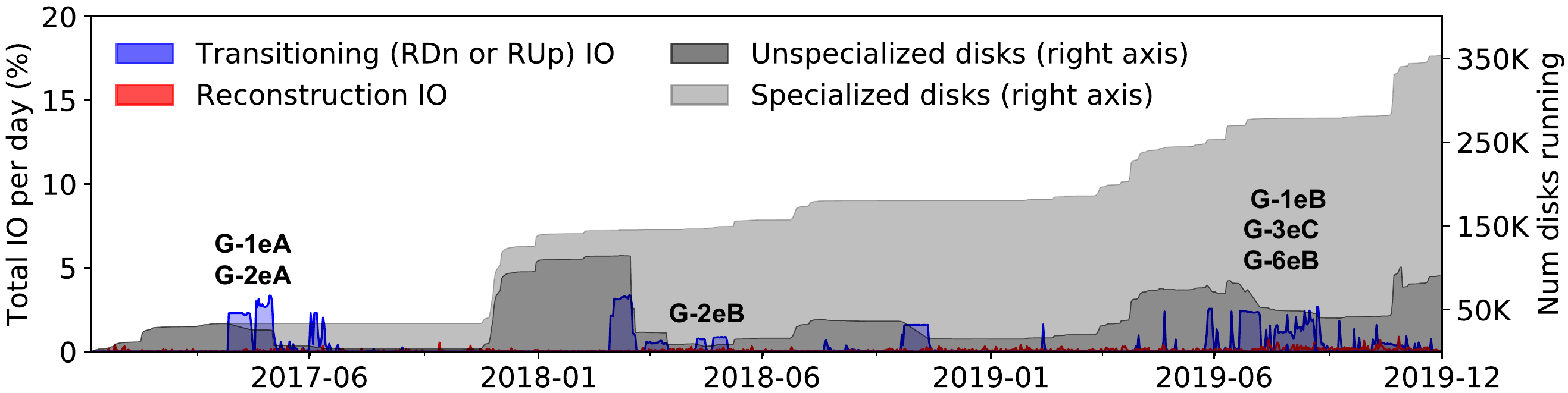}\vspace{-5pt}
        \caption{Redundancy management IO due to \pacemaker over its 2.5$+$ year lifetime broken down by IO type. This identical to Fig.~\ref{fig:pacer_front} with the left Y axis only going to 20\% to show the detailed IO activity happening in the cluster.}
        \label{fig:main_io_overheads}\vspace{-30pt}
    \end{subfigure}
    \begin{subfigure}[t]{0.22\textwidth}
        \includegraphics[width=\textwidth,right]{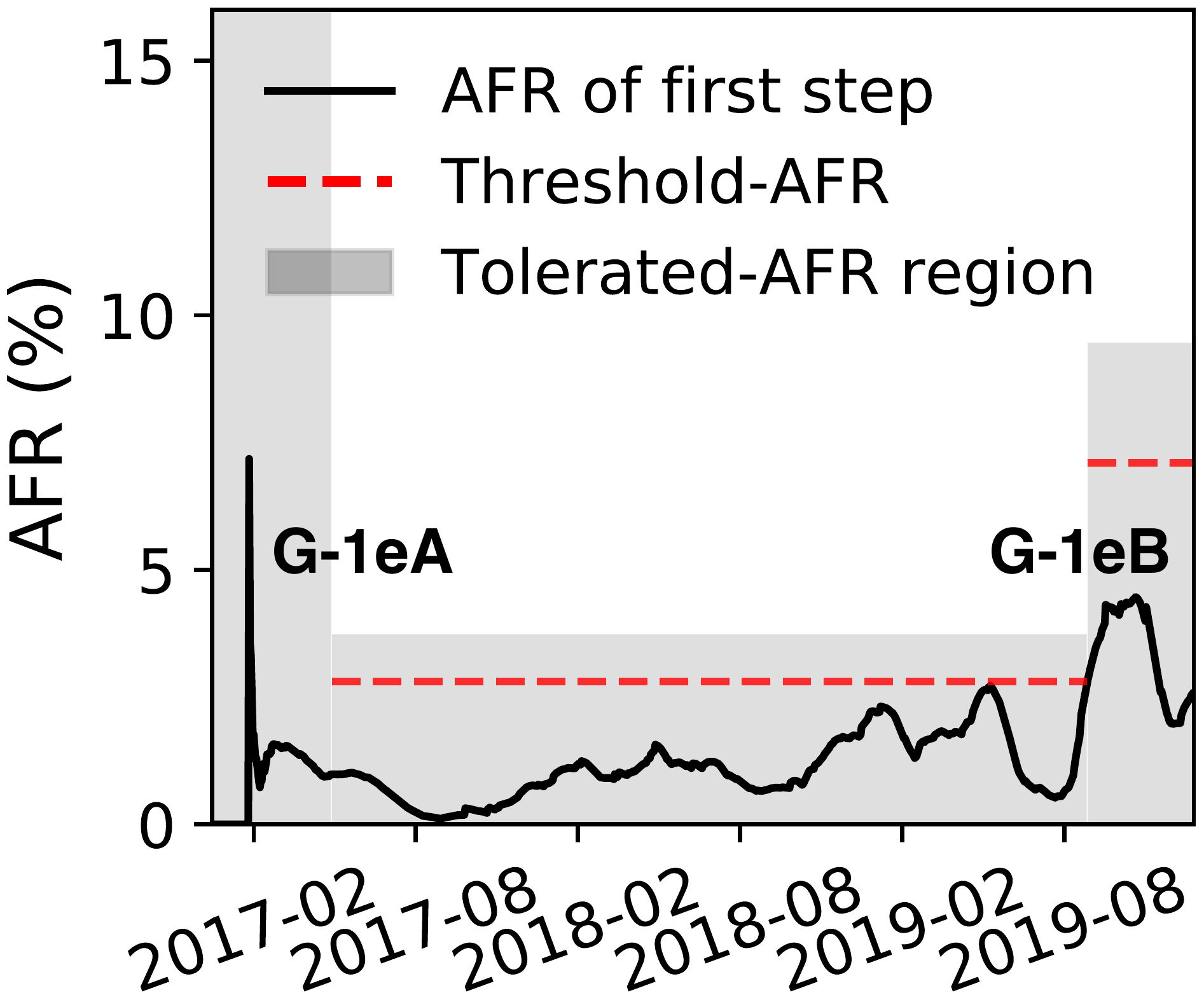}\vspace{-5pt}
        \caption{G-1 (step) AFR curve}
        \label{fig:main_afr_1}
    \end{subfigure}
    \begin{subfigure}[t]{0.755\textwidth}
        \hspace*{-0.46cm}
        \vspace{-5pt}
        \includegraphics[width=\textwidth,left]{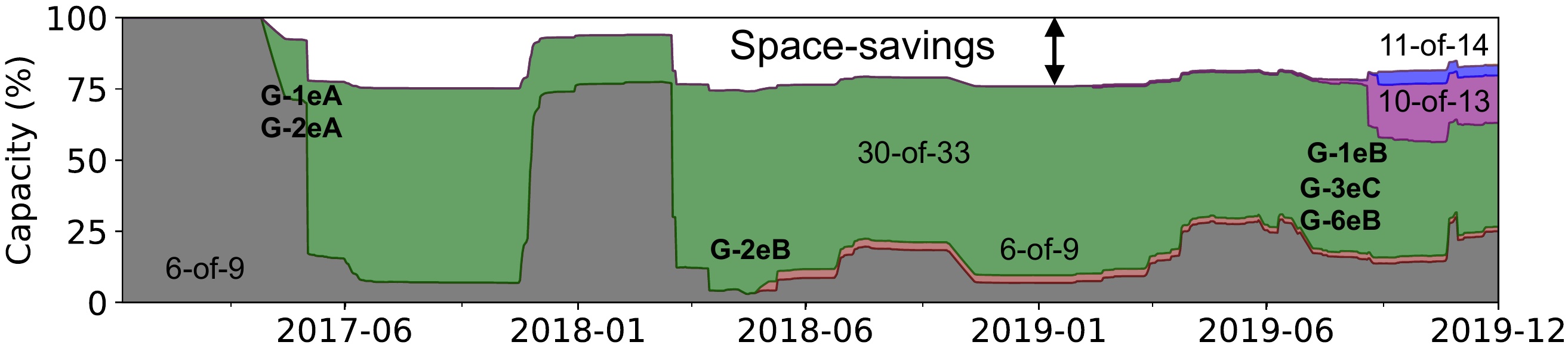}
        \caption{Space-savings due to \pacemaker. Each colored region represents the fraction of cluster capacity that is using a particular redundancy scheme. 6-of-9 is the default redundancy scheme (\argroup{0's}).}
        \label{fig:main_space_savings}
    \end{subfigure}
    \vspace{-3pt}
    \begin{subfigure}[t]{0.22\textwidth}
        \hspace*{0.08cm}
        \includegraphics[width=\textwidth,right]{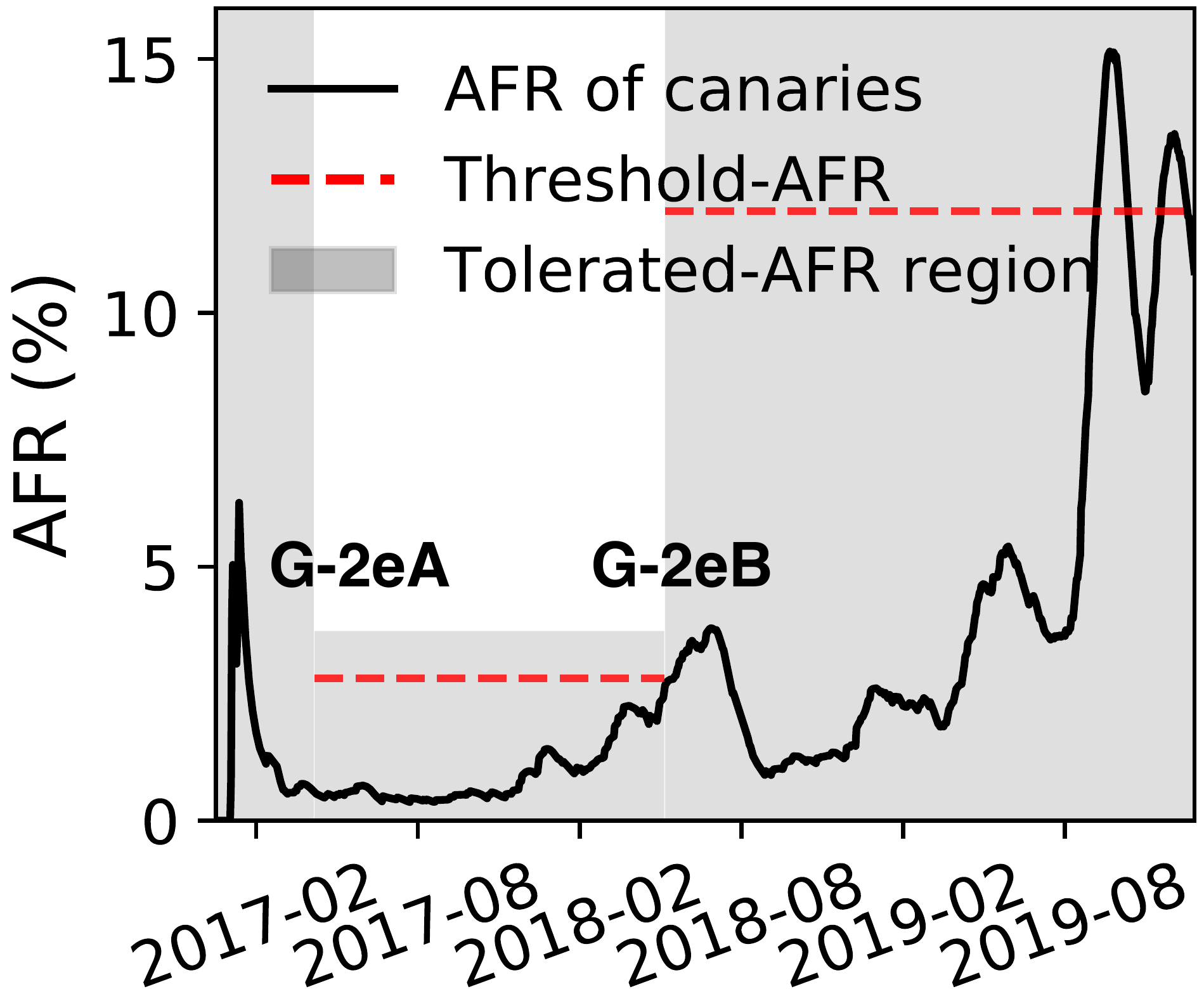}\vspace{-5pt}
        \caption{G-2 (trickle) AFR curve}
        \label{fig:main_afr_2}
    \end{subfigure}\vspace{-10pt}
    \caption{Detailed IO analysis and space savings achieved by \pacemaker{-enabled} adaptive redundancy on Google Cluster1. 
    \vspace{-15pt}}
    \label{fig:main_results}
\end{figure*}

\pacemaker does not change the file manipulation API or client access paths. 
But, there is one corner-case 
related to transitions when file reads can be affected internally. 
To read a file, a client queries the \namenode for the inode and caches it. Subsequently, the reads are performed directly from the client to the \datanode.
If the \datanode transitions to another \argroup while the file is still being read, the HDFS client may find that that \datanode no longer has the requested data. But, because this design uses existing HDFS decommissioning for transitions, the client software knows to react by re-requesting the updated inode from the \namenode and resuming the read.

%% file: sections/evaluation_v2.tex
\section{Evaluation} \label{sec:evaluation}
\pacemaker-enabled disk-adaptive redundancy using is evaluated on production logs from four large-scale real-world storage clusters, each with hundreds of thousands of disks. We also experiment with a proof-of-concept HDFS implementation on a smaller sized cluster. 
This evaluation has four primary takeaways:
(1) \pacemaker eliminates transition overload, never using more than 5\% of cluster IO bandwidth (0.2--0.4\% on average) and always meets target \mttdl, in stark contrast to prior work approaches that do not account for transition \io load;
(2) \pacemaker provides more than 97\% of idealized-potential \spacesavings, despite being proactive, reducing disk capacity needed by 14--20\% compared to one-size-fits-all;
(3) \pacemaker's behavior is not overly sensitive across a range of values for its configurable parameters;
(4) \pacemaker copes well with the real-world \afr characteristics explained in \S\ref{sec:production.afr}. For example, it successfully combines the
``multiple \usefullifephases{''} observation with  efficient transitioning schemes. 
This evaluation also shows \pacemaker in action by measuring \dadaptred in \pacemaker{-enhanced} HDFS.

\textbf{Evaluation methodology.}
\pacemaker is simulated chronologically for each of the four cluster logs described in \S\ref{sec:traceanalysis}: three clusters from Google and one from Backblaze.
For each simulated date, the simulator changes the cluster composition according to the disk additions, failures and decommissioning events in the log.
\pacemaker is provided the log information, as though it were being captured live in the cluster.
IO bandwidth needed for each day's redundancy management is computed as the sum of IO for failure reconstruction and transition IO requested by \pacemaker, and is reported as a fraction of the configured cluster IO bandwidth (100MB/sec per disk, by default).

\pacemaker was configured to use a \peakiocap of 5\%, an \avgio constraint of 1\% and a \thresholdafr of 75\% of the \toleratedafr, except for the sensitivity studies in \S\ref{sec:evaluation.sensitivity}.
For comparison, we also simulate (1) an idealized \dadaptred system in which transitions are instantaneous (requiring no IO) and (2) the prior state-of-the-art approach (\heart) for \dadaptred. 
For all cases, \argroup{0} uses 6-of-9, representing a one-size-fits-all scheme reported in prior literature~\cite{ford2010availability}. 
The required target \mttdl is then back-calculated using the 6-of-9 default and an assumed \toleratedafr of 16\% for \argroup{0}. \cradd{These configuration defaults were set by consulting storage \admins of clusters we evaluated. 
}

\subsection{\pacemaker on Google Cluster1 in-depth}
Fig.~\ref{fig:main_io_overheads} shows the IO generated by \pacemaker (and disk count) over the $\approx$3-year lifetime of Google Cluster1. Over time, the cluster grew to over \clusterthreesize disks comprising of disks from 7 makes/models (\adgroup{s}) via a mix of trickle and step deployments. 
Fig.~\ref{fig:main_afr_1} and Fig.~\ref{fig:main_afr_2} show \afr curves of 2 of the 7 \adgroup{s}\footnote{The rest of the \adgroup{s'} \afr curves are shown in Fig.~\ref{fig:main_results_1} in Appendix~\ref{app:remaining_clusters}.} (obfuscated as G-1 and G-2 for confidentiality) along with how \pacemaker adapted to them at each age. G-1 disks are \trickledeped whereas G-2 disks are \stepdeped. 
The other 5 \adgroup{s} are omitted due to lack of space.
Fig.~\ref{fig:main_space_savings} shows the corresponding \spacesavings~(the white space above the colors).

\begin{figure*}
    \begin{subfigure}[t]{0.33\textwidth}
        \includegraphics[width=\textwidth]{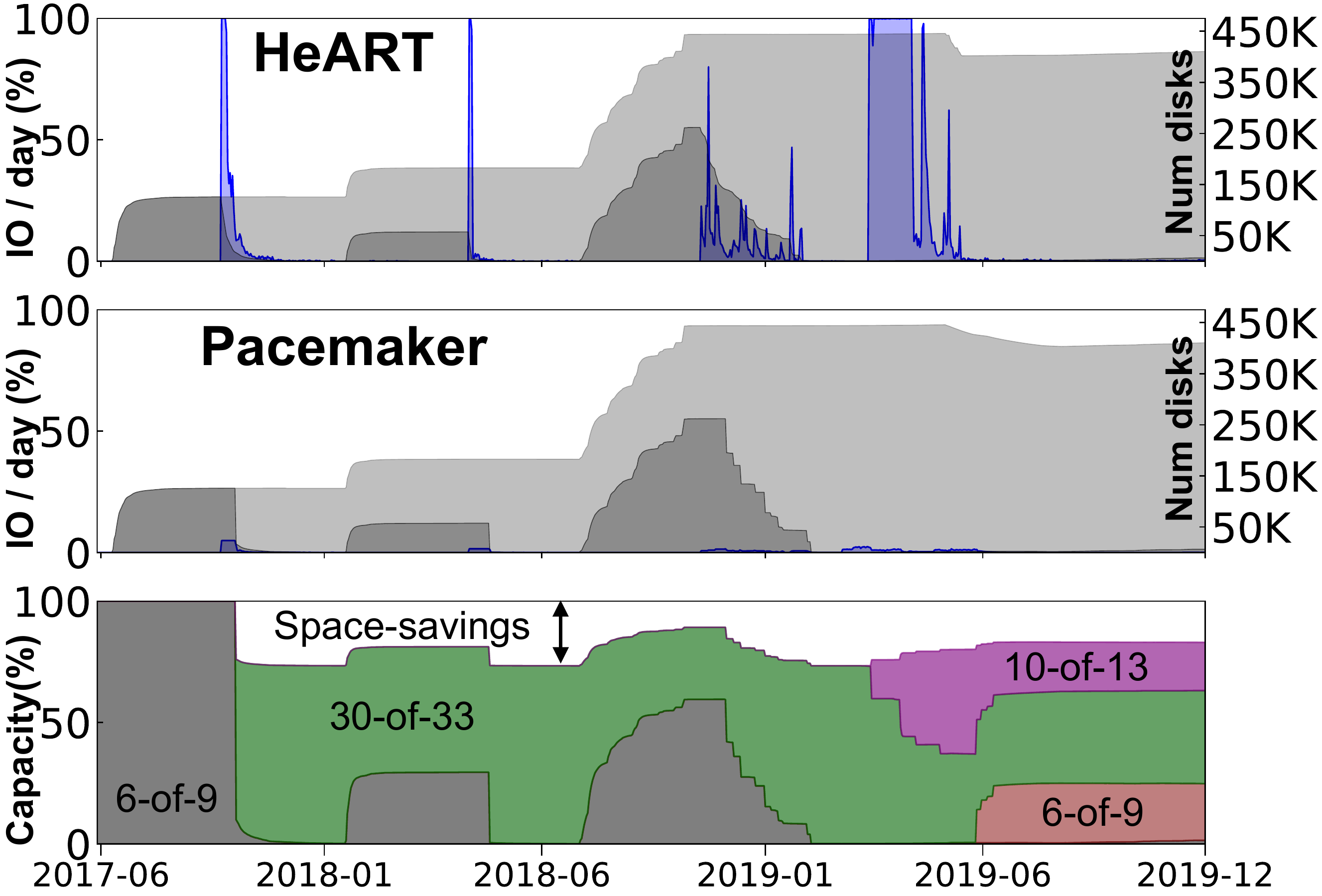}\vspace{-3pt}
        \caption{Google Cluster2}
        \label{fig:aux_cluster1_after}
    \end{subfigure}
    \begin{subfigure}[t]{0.33\textwidth}
        \includegraphics[width=\textwidth]{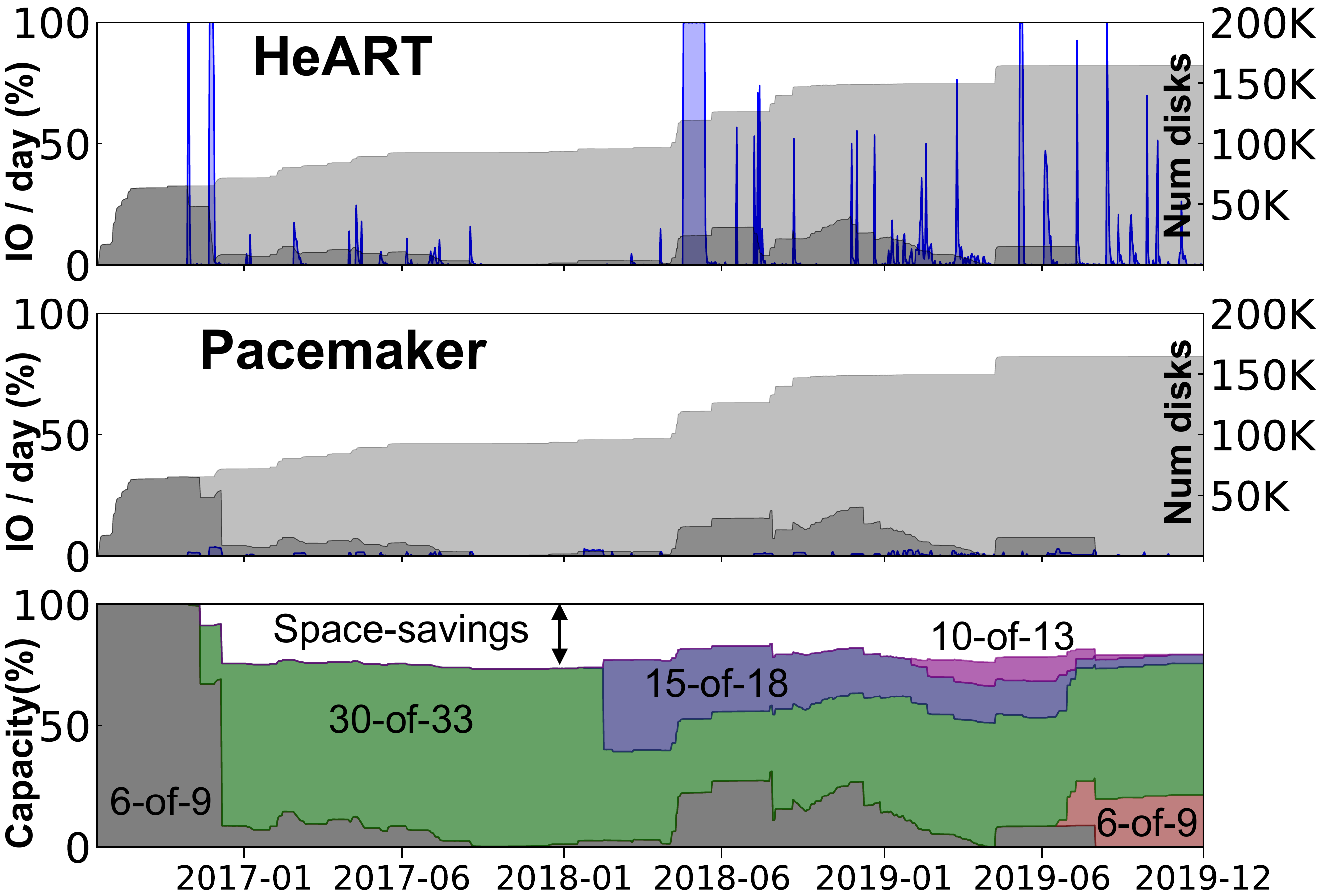}\vspace{-5pt}
        \caption{Google Cluster3}
        \label{fig:aux_cluster2_after}
    \end{subfigure}
    \begin{subfigure}[t]{0.33\textwidth}
        \includegraphics[width=\textwidth]{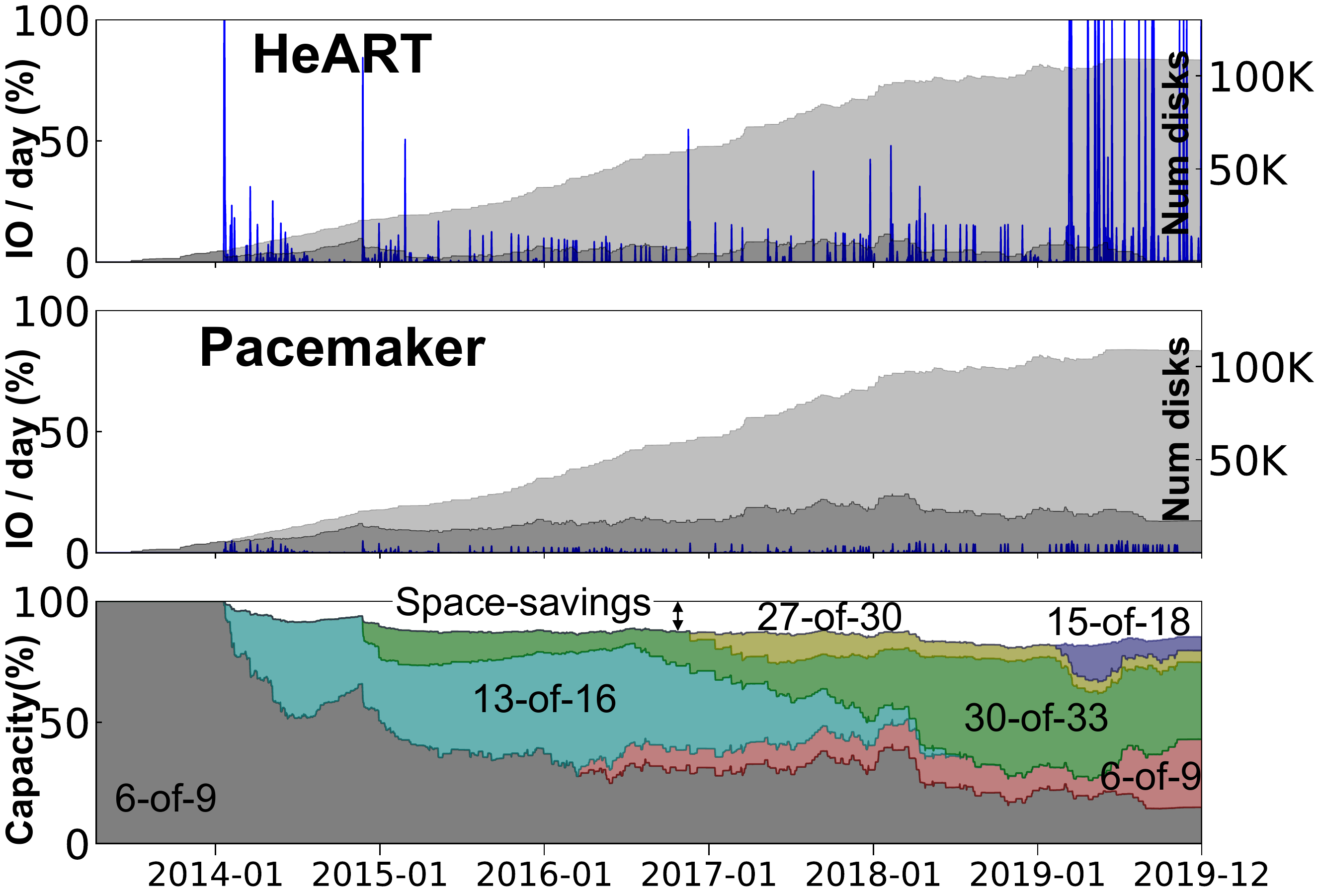}\vspace{-5pt}
        \caption{Backblaze}
        \label{fig:bb_after}
    \end{subfigure}
    \caption{Top two rows show the IO overhead comparison between prior adaptive redundancy system (\heart) and \pacemaker on two Google clusters and one Backblaze cluster. \pacemaker successfully bounds all IO under 5\% (visible as tiny blue regions in middle graphs, for e.g. around 2017 in \subref{fig:aux_cluster1_after}). The bottom row shows the 14--20\% average \spacesavings achieved by \pacemaker across the three clusters. The \afr curves of all three clusters are shown in Figs.~\ref{fig:main_results_2}--~\ref{fig:main_bb_results} in Appendix~\ref{app:remaining_clusters}.%
    \vspace{-15pt}
    }
    \label{fig:aux_clusters}
    \vspace{-1.em}
\end{figure*}

All disks enter the cluster as unspecialized disks, i.e. \argroup{0} (dark gray region in the Fig.~\ref{fig:main_io_overheads} and left gray region of Figs.~\ref{fig:main_afr_1} and~\ref{fig:main_afr_2}). 
Once a \adgroup{'s} \afr reduces sufficiently, \pacemaker \contraction{s} them to a specialized \argroup (light gray area in Fig.~\ref{fig:main_io_overheads}). Over their lifetime, disks may transition through multiple \expansion{s} over the multiple \usefullifephases. 
Each transition requires IO, which is captured in blue in Fig.~\ref{fig:main_io_overheads}. 
For example, the sudden drop in the unspecialized disks, and the blue area around 2018-04 captures the \transitionmethodtwo transitions caused when over 100K disks \contraction from \argroup{0} to a specialized \argroup. 
The light gray region in Fig.~\ref{fig:main_io_overheads} corresponds to the time over which \spacesavings are obtained, which can be seen in Fig.~\ref{fig:main_space_savings}.

\textbf{Many transitions with no transition overload.}
\pacemaker successfully bounds all redundancy management IO comfortably under the configured \peakiocap throughout the cluster's lifetime.
This can be seen via an imaginary horizontal line at 5\% (the configured \peakiocap) that none of the blue regions goes above.
Recall 
that \pacemaker 
\ratelimit{s} the IO within each \argroup to ensure simultaneous transitions do not violate the cluster's IO cap. 
Events \textbf{\textit{G-1eA}} and \textbf{\textit{G-2eA}} are examples of 
events where both G-1 and G-2 disks (making up almost 100\% of the cluster at that time) request transitions at the same time. Despite that, the IO remains bounded below 5\%. \textbf{\textit{G-3eC}} and \textbf{\textit{G-6eB}} 
also show huge disk populations of G-3 and G-6 \adgroup{s} 
(\afr{s} not shown) requesting almost simultaneous \expansion{s}, but \pacemaker's design ensures that the \peakio constraint is never violated. 
This is in sharp contrast with HeART's frequent transition overload, shown in Fig.~\ref{fig:heart_front}.

\textbf{Disks experience multiple \usefullifephases.}
G-1, G-3, G-6 and G-7 disks experience two phases of useful life each.
In Fig.~\ref{fig:main_io_overheads}, events \textbf{\textit{G-1eA}} and \textbf{\textit{G-1eB}} mark the two transitions of G-1 disks through its multiple useful lives as shown in Fig.~\ref{fig:main_afr_1}. In the absence of multiple \usefullifephases, \pacemaker would have \expansion{ed} G-1 disks to \argroup{0} in 2019-05, eliminating \spacesavings for the remainder of their time in the cluster.
\S\ref{sec:evaluation.ablation} quantifies the benefit of multiple \usefullifephases for all four clusters. 

\textbf{\mttdl always at or above target.}
Along with the \afr curves, Figs.~\ref{fig:main_afr_1} and~\ref{fig:main_afr_2} also show the upper bound on the \afr  
for which the reliability constraint is met (top of the gray region).
\pacemaker sufficiently protects all disks throughout their life for all \adgroup{s} across evaluated clusters.

\textbf{Substantial \spacesavings.}
\pacemaker provides 14\% average \spacesavings (Fig.~\ref{fig:main_space_savings}) over the cluster lifetime to date.
Except for 2017-01 to 2017-05 and 2017-11 to 2018-03, which correspond to infancy periods for large batches of new empty disks added to the cluster, the entire cluster achieves $\approx$20\% \spacesavings. 
Note that the apparent reduction in \spacesavings from 2017-11 to 2018-03 isn't actually reduced space in absolute terms. 
Since Fig.~\ref{fig:main_space_savings} shows relative \spacesavings, the over 100K disks deployed around 2017-11, and their infancy period makes the \spacesavings appear reduced relative to the size of the cluster. 

\begin{figure*}[t]
    \centering
    \begin{subfigure}[t]{0.53\textwidth}
        \includegraphics[width=\textwidth]{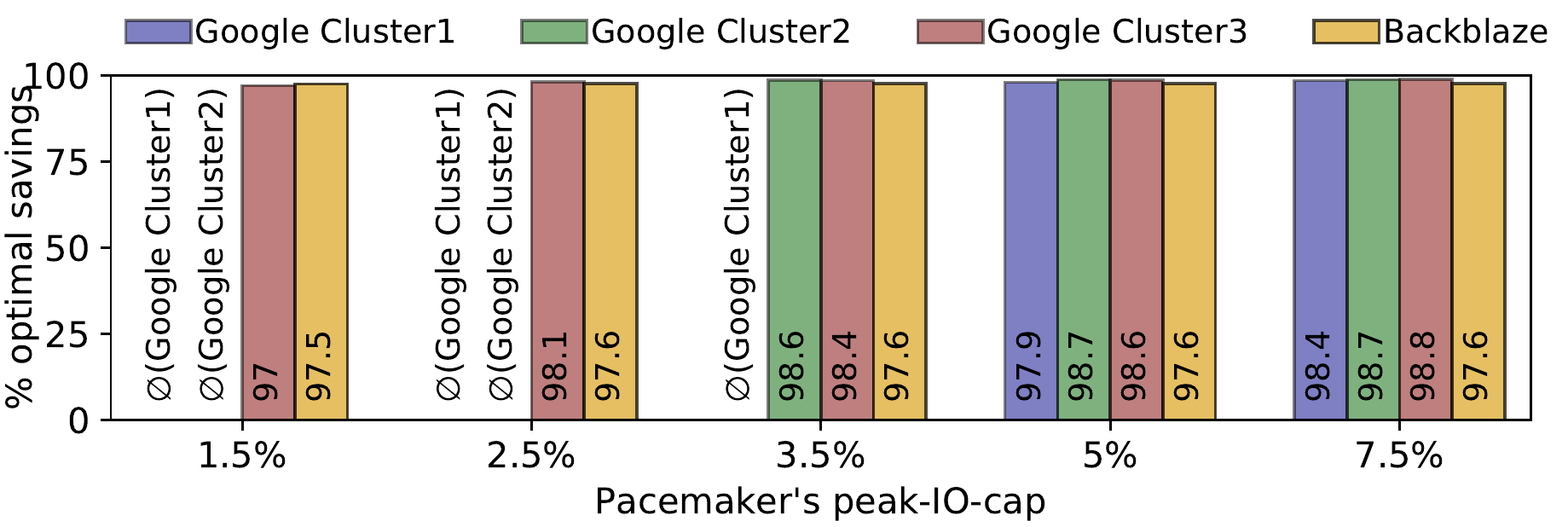}\vspace{-5pt}
        \caption{\pacemaker{'s} sensitivity to the \peakio constraint.}
        \label{fig:sensitivity_io}
    \end{subfigure}
    \begin{subfigure}[t]{0.23\textwidth}
        \centering
        \includegraphics[width=\textwidth]{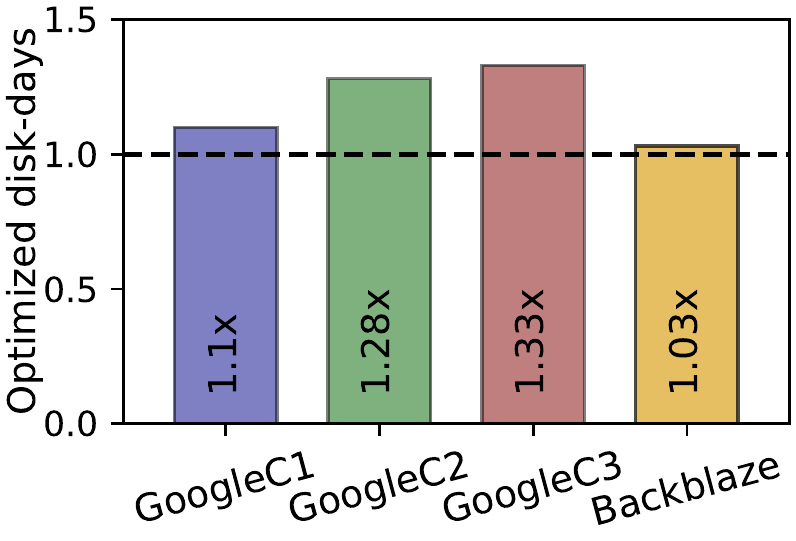}\vspace{-3pt}
        \caption{Multiple \usefullifephases}
        \label{fig:single_vs_multiple}
    \end{subfigure}
    \begin{subfigure}[t]{0.23\textwidth}
        \includegraphics[width=\textwidth]{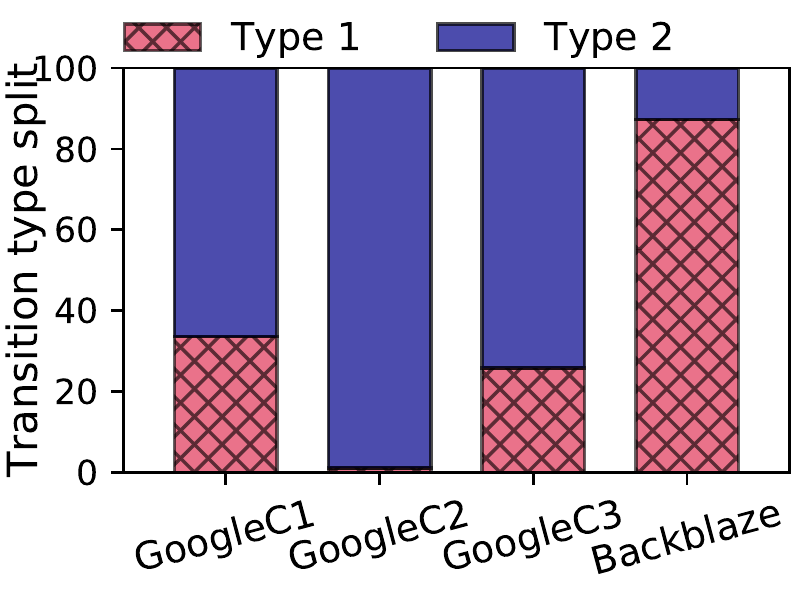}\vspace{-5pt}
        \caption{Transition type distribution}
        \label{fig:transition_type}
    \end{subfigure}
    \caption{\subref{fig:sensitivity_io} shows \pacemaker{'s} sensitivity to the peak IO bandwidth constraint. \subref{fig:single_vs_multiple} shows the advantage of multiple useful life phases and \subref{fig:transition_type} shows the contribution of the two transitioning techniques when \pacemaker was simulated on the four production clusters.%
    \vspace{-15pt}
    }\vspace{-8pt}
\end{figure*}

\subsection{\pacemaker on the other three clusters}
Fig.~\ref{fig:aux_clusters} compares the transition IO incurred by \pacemaker to that for \heart~\cite{kadekodi2019cluster} for Google Cluster2, Google Cluster3 and Backblaze, along with the corresponding \spacesavings achieved by \pacemaker.
While clusters using \heart would suffer transition overload, the same clusters under \pacemaker always had all their transition \io under the \peakiocap of 5\%. 
In fact, on average, only 0.21--0.32\% percent of the cluster IO bandwidth was used for transitions. The average \spacesavings{} for the three clusters are 14--20\%.

\cradd{
\textbf{Google Cluster2.} Fig.~\ref{fig:aux_cluster1_after} shows the \probname and \spacesavings in Google Cluster2 
and the corresponding \spacesavings. All \adgroup{s} in Google Cluster2 are \stepdeped. Thus, it is not surprising that Fig.~\ref{fig:transition_type} shows that over 98\% of the transitions in Cluster2 were \transitionmethodtwo transitions (bulk parity recalculation). 
Cluster2's disk population exceeds 450K disks. Even at such large scales, \pacemaker obtains average \spacesavings of almost 17\% and peak \spacesavings of over 25\%. This translates to needing 100K fewer disks.

\textbf{Google Cluster3.} Google Cluster3 (Fig.~\ref{fig:aux_cluster2_after}) is not as large as Cluster1 or Cluster2. At its peak, Cluster3 has a disk population of approximately 200K disks. But, it achieves the highest average \spacesavings (20\%) among clusters evaluated. Like Cluster2, Cluster3 is also mostly \stepdeped.

\textbf{Backblaze Cluster.} Backblaze (Fig.~\ref{fig:bb_after}) is a completely \trickledeped cluster. The dark grey region across the bottom of Fig.~\ref{fig:bb_after}'s \pacemaker plot shows the persistent presence of canary disks throughout the cluster's lifetime. Unlike the Google clusters, the transition IO of Backblaze does not produce bursts of transition \io that lasts for weeks. Instead, since \trickledeped disks transition a-few-at-a-time, we see transition work appearing continuously throughout the cluster lifetime of over 6 years. The rise in the transition IO spikes in 2019, for \heart, is because of large capacity 12TB disks replacing 4TB disks. Unsurprisingly, under \pacemaker, most of the transitions are done using \transitionmethodone (transitioning by emptying disks) as shown in Fig.~\ref{fig:transition_type}. The average \spacesavings obtained on Backblaze are 14\%.} 

\subsection{Sensitivity analyses and ablation studies} 
\label{sec:evaluation.ablation}
\label{sec:evaluation.sensitivity}

\textbf{Sensitivity to IO constraints.} The \peakio constraint governs 
Fig.~\ref{fig:sensitivity_io}, which shows the percentages of optimal \spacesavings achieved with \pacemaker for 
\peakiocap settings between 1.5\% and 7.5\%. \cradd{A \peakiocap of up to 7.5\% is used in order to compare with the \io percentage spent for existing background IO activity, such as scrubbing. By scrubbing all data once every 15 days~\cite{bairavasundaram2007analysis}, the scrubber uses around 7\% IO bandwidth, and is a background work \io level tolerated by today's clusters.}

The Y-axis captures how close the \spacesavings are for the different \peakiocap{s} compared to ``Optimal savings'', 
i.e. an idealized system with infinitely fast transitions. 
\pacemaker's default \peakiocap (5\%) achieves over 97\% of the optimal \spacesavings for each of the four clusters.
For \peakio constraint set to $<=$2.5\%, some \expansion{s} in Google Cluster1 and Cluster2 
become too aggressively \ratelimit{ed} causing a subsequent \afr rise to violate the \peakio constraints. 
We indicate this as a failure, and show it as "$\varnothing$".
The same situation happens for Google Cluster1 at 3.5\%. 

\textbf{Sensitivity to \thresholdafr.} 
The \thresholdafr determines when proactive \expansion{s} of \stepdeped disks are initiated. 
Conceptually, the \thresholdafr governs how risk-averse the admin wants to be.
Lowering the threshold would trigger an \expansion when disks are farther away from the \toleratedafr (more risk-averse), and vice-versa. 
We evaluated \pacemaker for \thresholdafr{s} of 60\%, 75\% and 90\% of the respective \argroup{s'} \toleratedafr{s}.
We found that \pacemaker's \spacesavings is not very sensitive to \thresholdafr, with \spacesavings only 2\% lower at 60\% than at 90\%.
Data remained safe at each of these settings, but would become unsafe with higher values.

\textbf{Contribution of multiple \usefullifephases.} Fig.~\ref{fig:single_vs_multiple} compares the increased number of disk-days spent in specialized \argroup{s} because of considering multiple \usefullifephases. In the best case, Google Cluster2 spent 33\% more disk-days in specialized redundancy, increasing overall \spacesavings from 16\% to 19\%. 
Note that in large-scale storage clusters, even 1\% \spacesavings are considered substantial
as it represents thousands of disks.

\textbf{Contribution of transition types.}
By proactively keeping \stepdeped disks in distinct \argroups and using specialized transitioning schemes whenever possible, instead of using simple re-encoding for all transitions, \pacemaker reduces total transition IO by 92--96\% for the four clusters.
Fig.~\ref{fig:transition_type} shows what percentage of transitions were done via \transitionmethodone (disk emptying) vs. \transitionmethodtwo (bulk parity recalculation). 
As expected, Google clusters rely more on \transitionmethodtwo transitions, because most disks are \stepdeped.
In contrast, the Backblaze cluster is entirely \trickledeped and hence mostly uses \transitionmethodone transitions. 
The small percentage of \transitionmethodtwo transitions 
in Backblaze occur when \argroup{s} are purged. 

\subsection{Evaluating HDFS + \pacemaker}
\label{sec:hdfs_eval}
This section describes basic experiments with the \pacemaker{-enabled} HDFS, focusing on its functioning and operation. Note that \pacemaker is designed for longitudinal disk deployments over several years, a scenario that cannot be reproduced identically in laboratory settings. Hence, these HDFS experiments are aimed to display that integrating \pacemaker with an existing storage system is straightforward, rather than on the long-term aspects like overall \spacesavings or transition \io behavior over cluster lifetime as evaluated via simulation above.

The HDFS experiments run on a PRObE Emulab cluster~\cite{gibson2013probe}.
Each machine has a Dual-Core AMD Opteron Processor, 16GB RAM, and Gigabit Ethernet.
We use a 21-node cluster running HDFS 3.2.0 with one \namenode and 20~\datanode{s}. \cradd{Each \datanode has a 10GB partition on a 10000 RPM HDD for a total cluster size of 200GB.}
We statically define the cluster to be made up of two \argroup{s} of ten \datanode{s} each, one using the 6-of-9 erasure coding scheme and the other using a 7-of-10 scheme.
DFS-perf~\cite{gu2016dfs}, a popular open-source HDFS benchmark is used, after populating the cluster to 60\% full.
Each DFS-perf client sequentially reads one file over and over again (size=768MB), for a total read size of about 1.75TB over 40 iterations.
We use 60 DFS-perf clients, running on 20~nodes separate from the HDFS cluster.

\begin{figure}[t]
\centering
\includegraphics[width=0.95\textwidth]{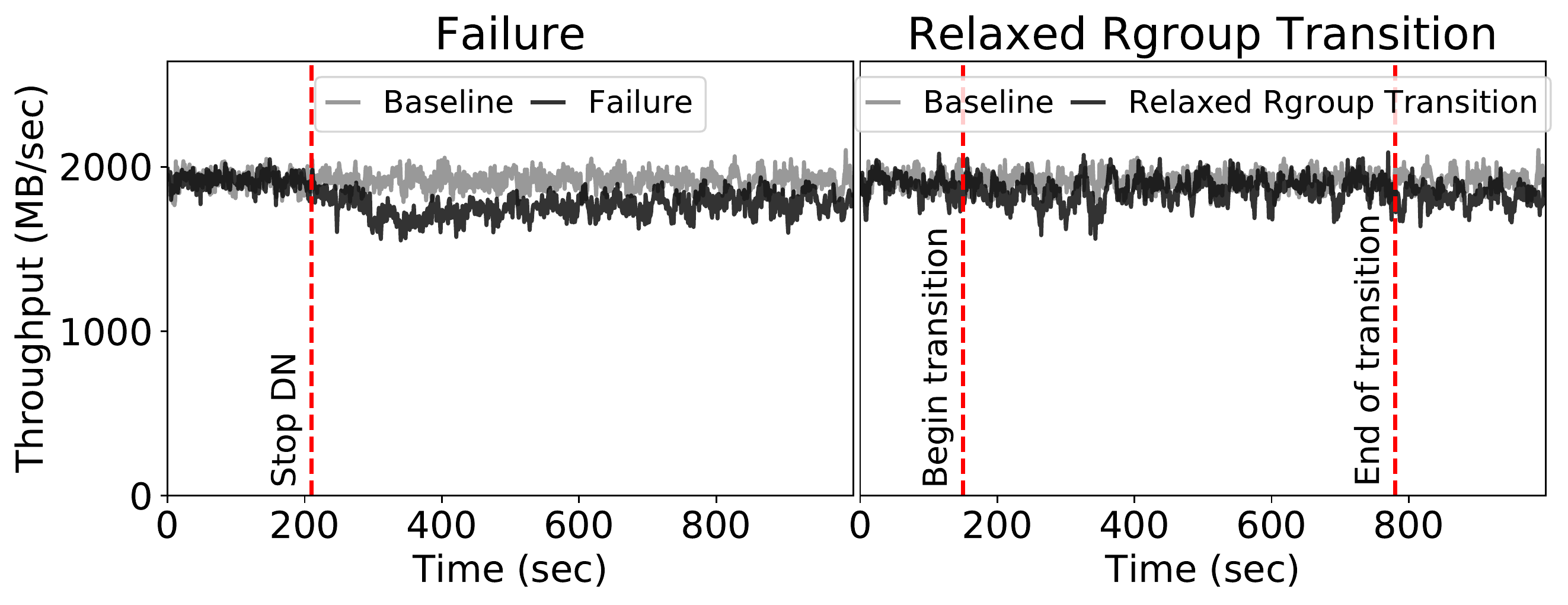}\vspace{-5pt}
\caption{DFS-perf reported throughput for baseline, with one \datanode failure and one \argroup transition.%
\vspace{-5pt}}
\label{fig:hdfs_throughput}
\end{figure}

We focus on the behavior of a \datanode as it transitions between \argroup{s},  compared with baseline HDFS performance (where all \datanode{s} are healthy) and its behavior while recovering from a failed \datanode. Fig.~\ref{fig:hdfs_throughput} shows the client throughput after the setup phase,
followed by a noticeable drop in client throughput when a \datanode fails (emuated by stopping the DN). This is caused by the reconstruction IO that recreates the data from the failed node. Read latency 
exhibits similar behavior (not shown due to space). Eventually, throughput settles at about 5\% lower than prior to failure, since now there are 19~\datanode{s}.

Fig.~\ref{fig:hdfs_throughput} also shows client throughput when a node is \contracted from 
6-of-9 to 7-of-10. 
There is minor interference during the transition, which can be attributed to the data movement that HDFS performs as a part of decommissioning. The transition requires less work than failed node reconstruction, yet takes longer to complete because \pacemaker limits the \tio.
Eventually, even though 20 \datanode{s} are running, the throughput is lower by $\approx$5\% (one \datanode{'s} throughput). This happens because \pacemaker empties the \datanode before it moves into the new \argroup, and load-balancing data to newly added \datanode{s} happens over a longer time-frame. Experiments with \expansion{} showed similar results.

%% file: sections/related_work.tex
\section{Related work}
\label{sec:related_work}
The closest related work~\cite{kadekodi2019cluster} proposes a redundancy adaptation tool called \heart that categorizes disks into groups and suggests a tailored redundancy scheme for each during its useful life period. 
As discussed earlier, while~\cite{kadekodi2019cluster} 
showcased potential \spacesavings, it ignored \probname and hence is made impractical (Fig.~\ref{fig:heart_front}).
\pacemaker eliminates \probname 
by employing \io constraints (specifically the \peakio and \avgio constraints) that cap the transition IO to a tiny fraction cluster bandwidth. 
While \heart was evaluated only for the \trickledeped Backblaze cluster,
our evaluation of \pacemaker{} for Google storage clusters exposes the unique challenges of \stepdeped clusters. Several design elements were added to \pacemaker to address the challenges posed by \stepdeped disks.

Various systems include support for multiple redundancy schemes, allowing different schemes to be used for different data~\cite{hdfs-ec,ceph-ec}.
Tools have been created for deciding, on a per-data basis, which scheme to use~\cite{Xia2015,thereska2006informed}.
Keeton et al.~\cite{keeton2004designing} describe a tool that automatically provides disaster-resistant solutions based on budget and failure models.
\pacemaker differs from such systems by focusing on efficiently adapting redundancy to different and time-varying \afr{s} of 
disks.

Reducing the impact of background IO, such as for data scrubbing, on foreground IO is a common research 
theme.~\cite{amvrosiadis2015opportunistic, bachmat2002analysis, lumb2000towards, lumb2002freeblock, oprea2010clean, schwarz2004disk}.
\pacemaker 
converts otherwise-urgent bursts of transition IO into proactive background IO, which could then benefit from these works. 

Disk reliability has been well studied, including evidence of failure rates being make/model dependent~\cite{bairavasundaram2007analysis, elerath2009hard, heien2011modeling, jiang2008disks, ma2015raidshield, patterson1988case, pinheiro2007failure, schroeder2010understanding, schroeder2007disk, schroeder2007understanding, shah2004disk}.
There are also studies that predict disk failures~\cite{anantharaman2018large, hamerly2001bayesian, mahdisoltani2017proactive, murray2003hard, strom2007hard, wang2014two, zhao2010predicting}, \nocite{chun2006efficient, sit2006proactive, thereska2006informed}
which can enhance any storage fault-tolerance approach.

While several works have considered the problem of designing erasure codes that allow transitions using less resources, existing solutions are limited to specific kinds of transitions and hence are not applicable in general. The case of adding parity chunks while keeping the number of data chunks fixed can be viewed~\cite{rashmi2011enabling,rashmi2017piggybacking, mousavi2018delayed} as the well-studied reconstruction problem, and hence the codes designed for optimal reconstruction (e.g.,~\cite{dimakis2010network,rashmi2011optimal,rashmi2017piggybacking, vajha2018clay,gopalan2012locality,papailiopoulos2014locally}) would lead to improved resource usage for this case. Several works have studied the case where the number of data nodes increases while the number of parity nodes remains fixed \cite{zheng2011fastscale,wu2016i/o,zhang2018optimal,hu2018generalized,rai2015adaptivea}. In \cite{Xia2015}, the authors propose two erasure codes designed to undergo a specific transition in parameters. In \cite{maturana2020convertible}, the authors propose a general theoretical framework for studying codes that enable efficient transitions for general parameters, and derive lower bounds on the cost of transitions as well as describe optimal code constructions for certain specific parameters. However, none of the existing code constructions are applicable for the diverse set of transitions needed for disk-adaptive redundancy in real-world storage clusters.

%% file: sections/conclusion.tex
\section{Conclusion}
\label{sec:conclusion}

\pacemaker orchestrates disk-adaptive redundancy without \probname{}, allowing use in real-world clusters. By proactively arranging data layouts and initiating 
transitions, \pacemaker reduces total transition IO allowing it to be rate-limited.
Its design integrates cleanly into existing scalable storage implementations, such as HDFS.
Analysis for 4 large real-world storage clusters from Google and Backblaze show 14--20\%  average space-savings while transition IO is kept small ($<$0.4\% on average) and bounded (e.g., $<$5\%).

%% file: sections/ack.tex
\section{Acknowledgements}
We thank our shepherd Wyatt Lloyd and the anonymous reviewers for their valuable feedback and suggestions. We extend special thanks to Larry Greenfield, Arif Merchant and numerous other researchers, engineers at Google; Keith Smith, Tim Emami, Jason Hennessey, Peter Macko and other researchers from NetApp's Advanced Technology Group (ATG) who have been instrumental in providing data, feedback and support. We also thank Jiaan Dai, Xuren Zhou, Jiaqi Zuo, Sai Kiriti Badam and Jiongtao Ye for their help in building the HDFS+\pacemaker prototype. This research is generously supported in part by the NSF grants CNS 1956271 and CNS 1901410. We also thank the members and companies of the PDL Consortium (Alibaba,
Amazon,
Datrium,
Facebook,
Google,
HPE,
Hitachi,
IBM,
Intel,
Microsoft,
NetApp,
Oracle,
Pure Storage,
Salesforce,
Samsung,
Seagate,
Two Sigma, Western Digital) and VMware for their interest, insights, feedback, and support.

%% file: sections/new_appendix.tex
\section{Failure rate estimation details}
\label{app:afr_estimation}
This section describes how we calculate failure rates for each \adgroup\ based on the disks' age using empirical data.
In the storage device reliability literature, the failure rate over a period of time is typically expressed in terms of Annualized Failure Rate (AFR), and calculated as:
\begin{equation}\label{eq:afr_estimate}
    AFR\ (\%) = \frac{d}{E} \times 100,
\end{equation}
where $d$ is the number of observed disk failures, and $E$ is the sum of the exposure time of each disk, measured in years.
The exposure time of a disk is the amount of time it was in operation (i.e., deployed and had not failed nor been retired) during the period in consideration, and it is typically measured at the granularity of days.

If the time to failure is exponentially distributed, then Formula~\ref{eq:afr_estimate} corresponds to the maximum likelihood estimate for the rate parameter of the exponential distribution.
Due to the memoryless property of this distribution, such a formula would be appropriate only if we assume that failure rate is constant with respect to time or device age.
Thus, Eq.~\ref{eq:afr_estimate} may be useful for estimating AFR over long and stable periods of time, but makes it hard to reason about changes in AFR over time.
Therefore, in this work, we estimate AFR using the following approach.

Assume that the lifetime (time from deployment to failure) of each disk is an i.i.d.~discrete random variable $T$ with cumulative density function $F$ and probability mass function $f$. 
The failure rate (also known as \textit{hazard rate}) \cite{trivedi2002probability} of this distribution is given by:
\begin{equation}
    h(t) = f(t) / (1 - F(t)).
\end{equation}
The \emph{cumulative hazard} defined as $H(t) = \sum_{i=0}^t h(i)$ is commonly estimated using the Nelson-Aalen estimator:
\begin{equation}\label{eq:cumhaz}
    \hat{H}(t) = \sum_{i=0}^t \frac{d_i}{a_i}\qquad \text{for } t \in \{0, \ldots, m\},
\end{equation}
where $d_i$ is the number of disks that failed during their $i$-th day, $a_i$ is the number of disks that were in operation at the start of their $i$-th day, and $m$ is the age in days of the oldest observed disk drive.
An estimate for the failure rate can be obtained by applying the so-called \emph{kernel method}~\cite{tanner1983}:
\begin{equation}\label{eq:haz}
    \hat{h}(t) = \sum_{i=0}^{m} \frac{d_i}{a_i} K(t - i),\qquad \text{for } t \in \{0, \ldots, m\},
\end{equation}
where $K(\cdot)$ is a kernel function.
In practice, Formula~\ref{eq:haz} can be considered as a smoothing over the increments of Formula~\ref{eq:cumhaz}.
For our calculations, we utilized an Epanechnikov kernel~\cite{hastie2009kernel} with a bandwidth of 30 days (the Epanechnikov kernel is frequently used in practice due to its good theoretical properties).

A big advantage of this approach is that it is \emph{nonparametric}, meaning that it does not assume that the lifetime $T$ follows any particular distribution.
This allows \pacemaker\ to adapt and work effectively with a wide arrange of storage devices with vastly different failure rate behaviors.

\section{Detailed cluster evaluations} \label{app:remaining_clusters}
\begin{figure*}[t]
    \centering
    \begin{subfigure}[t]{0.24\textwidth}
        \centering
        \includegraphics[width=\textwidth]{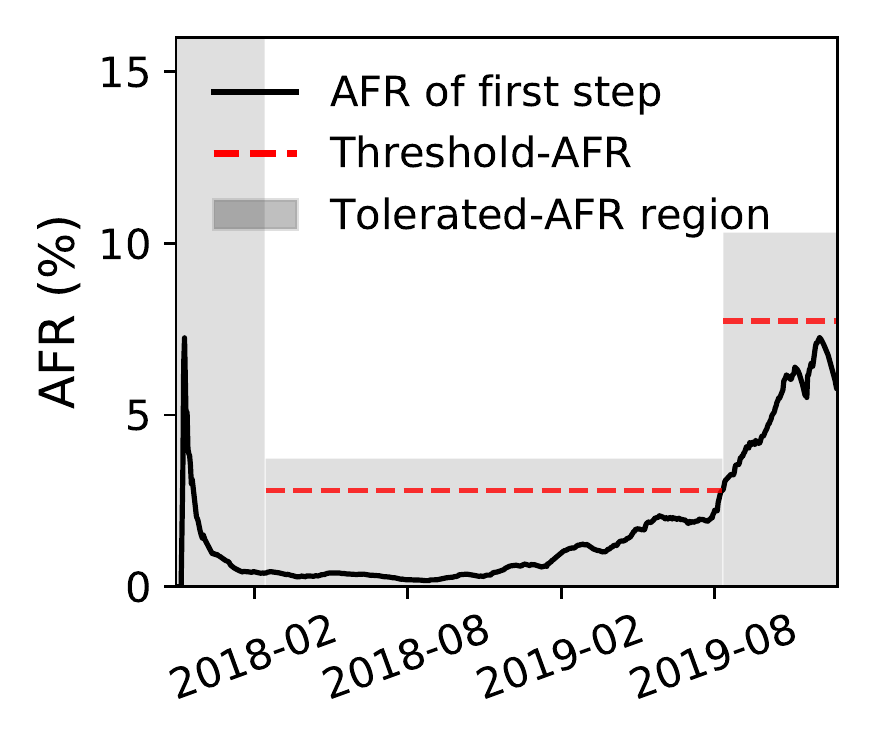}\vspace{-5pt}
        \caption{G-6 (step) \afr curve}
        \label{fig:g_cluster3_afr3}
    \end{subfigure}
    \begin{subfigure}[t]{0.24\textwidth}
        \centering
        \includegraphics[width=\textwidth]{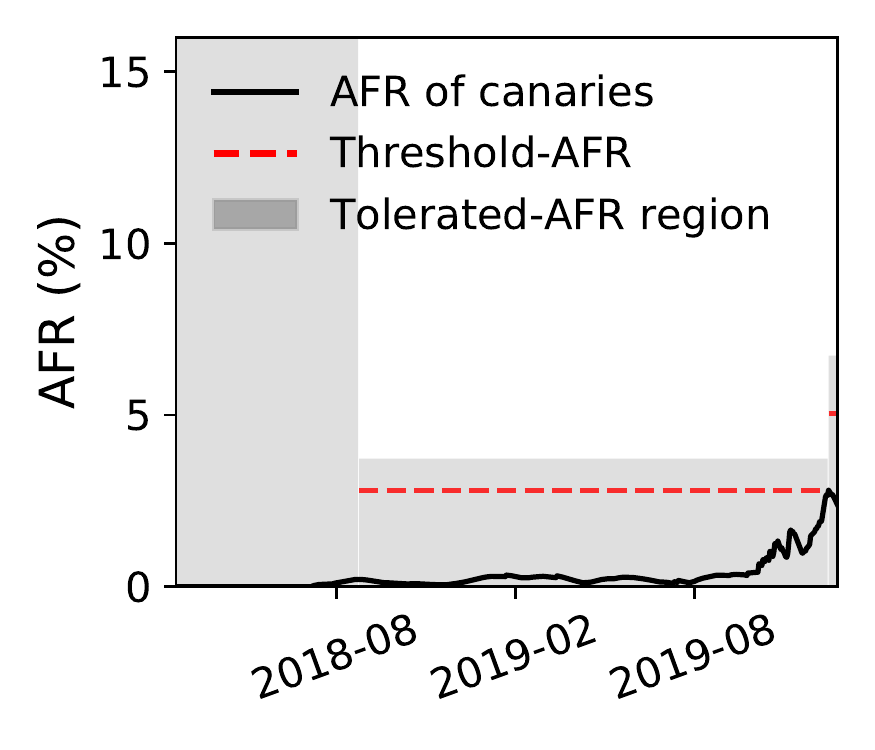}\vspace{-5pt}
        \caption{G-4 (trickle) \afr curve}
        \label{fig:g_cluster3_afr4}
    \end{subfigure}
    \begin{subfigure}[t]{0.24\textwidth}
        \centering
        \includegraphics[width=\textwidth]{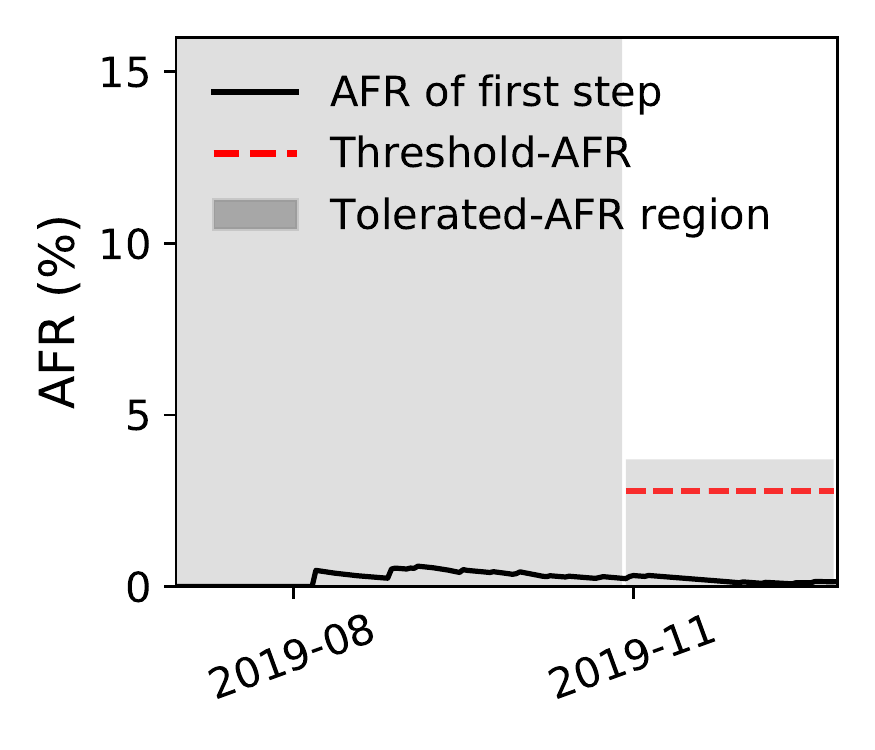}\vspace{-5pt}
        \caption{G-8 (step) \afr curve}
        \label{fig:g_cluster3_afr5}
    \end{subfigure}
    \begin{subfigure}[t]{0.24\textwidth}
        \centering
        \includegraphics[width=\textwidth]{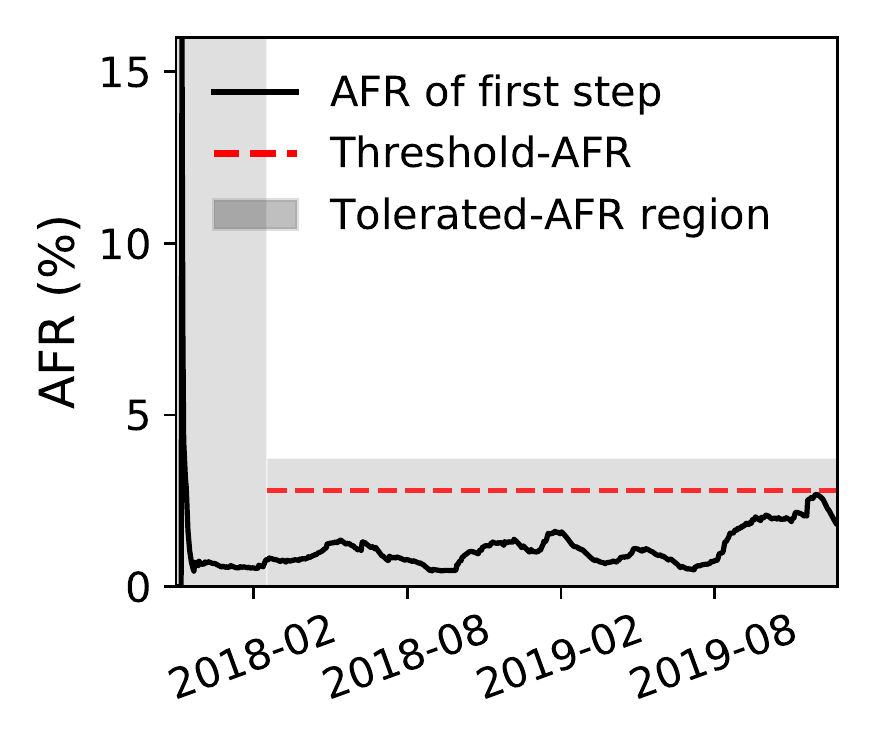}\vspace{-5pt}
        \caption{G-5 (step) \afr curve}
        \label{fig:g_cluster3_afr6}
    \end{subfigure}
    \begin{subfigure}[t]{0.24\textwidth}
        \centering
        \includegraphics[width=\textwidth]{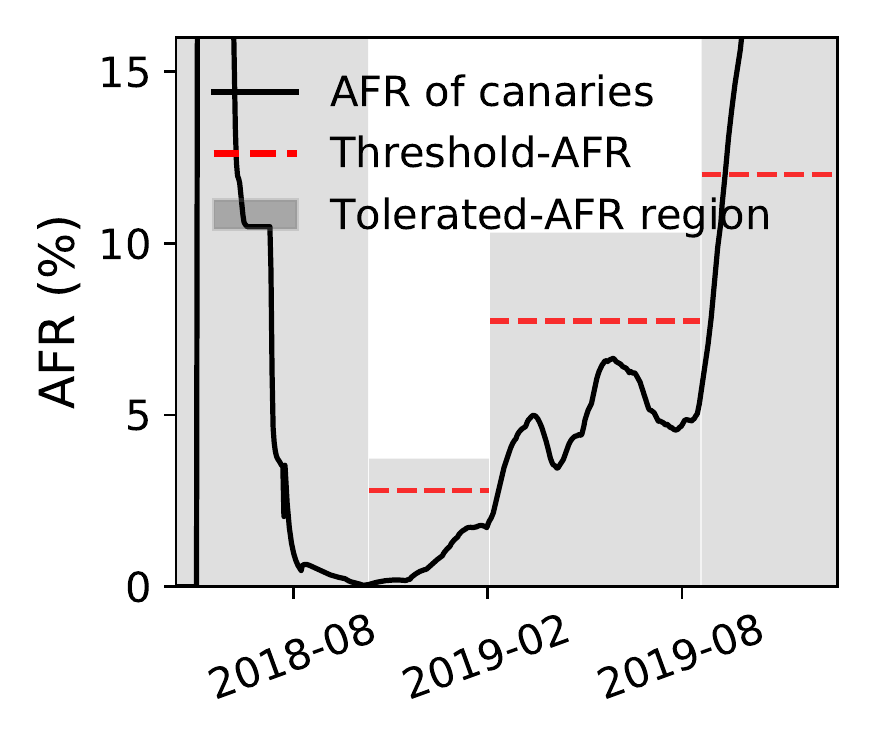}\vspace{-5pt}
        \caption{G-3 (trickle) \afr curve}
        \label{fig:g_cluster3_afr7}
    \end{subfigure}
    \caption{Detailed IO analysis and space savings achieved by \pacemaker{-enabled} adaptive redundancy on Google Cluster1.}
    \label{fig:main_results_1}
\end{figure*}

This section shows the remaining \adgroup{s} of Google Cluster1 (Fig.~\ref{fig:main_results_1}) and provides a similar deep-dive of \pacemaker on Google Cluster2, Google Cluster3 and the Backblaze cluster along with the \afr curves of all \adgroup{s} of those clusters. 

\textbf{\adgroup{s} of Google Cluster1.} Recall from \S\ref{sec:evaluation} that Google Cluster1 is made up of seven \adgroup{s}. G-1 and G-2 \afr curves are shown in Figs.~\ref{fig:main_afr_1} and~\ref{fig:main_afr_2} respectively. Here we show the four of the file remaining \adgroup{s}, viz. G-3, G-5, G-6 and G-7 in Figs.~\ref{fig:g_cluster3_afr3}--\ref{fig:g_cluster3_afr7}. G-3 and G-7 disks are \trickledeped similar to G-2 disks, whereas the other disks are \stepdeped. 

\begin{figure*}[b]
    \centering
    \begin{subfigure}[t]{0.8\textwidth}
        \includegraphics[width=\textwidth]{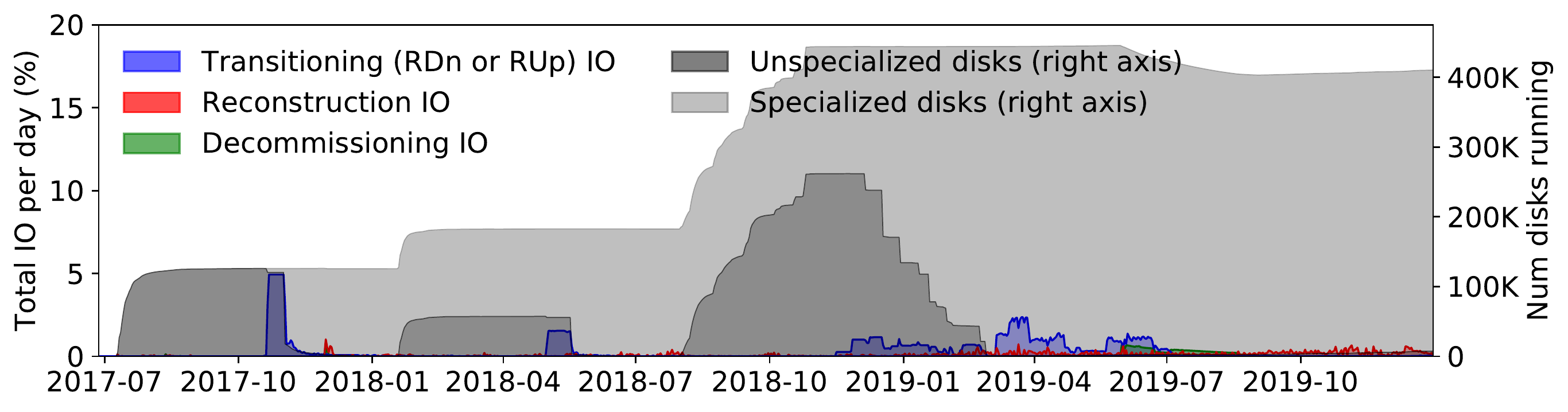}\vspace{-5pt}
        \caption{Google Cluster2 redundancy management IO due to \pacemaker over its 2$+$ year lifetime broken down by IO type.}
        \label{fig:g_cluster1_after_20}
    \end{subfigure}
    \begin{subfigure}[t]{0.76\textwidth}
        \hspace*{-0.85cm}
        \includegraphics[width=\textwidth]{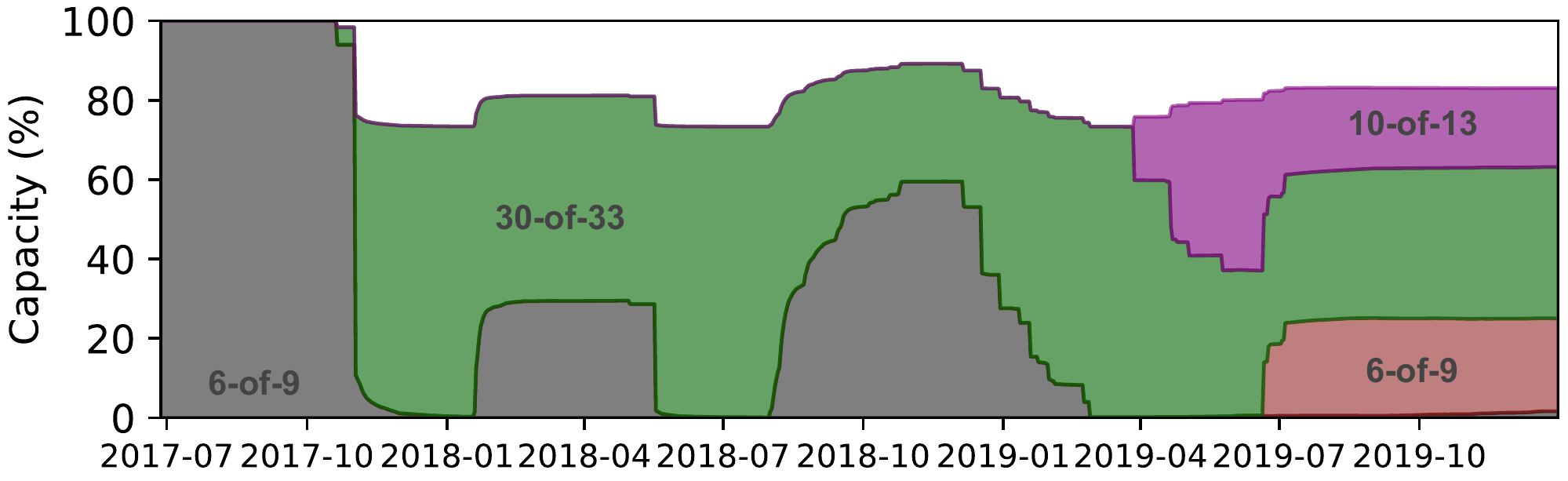}\vspace{-5pt}
        \caption{Google Cluster2 space savings achieved by \pacemaker.}
        \label{fig:g_cluster1_space}
    \end{subfigure}
    \begin{subfigure}[t]{0.24\textwidth}
        \centering
        \includegraphics[width=\textwidth]{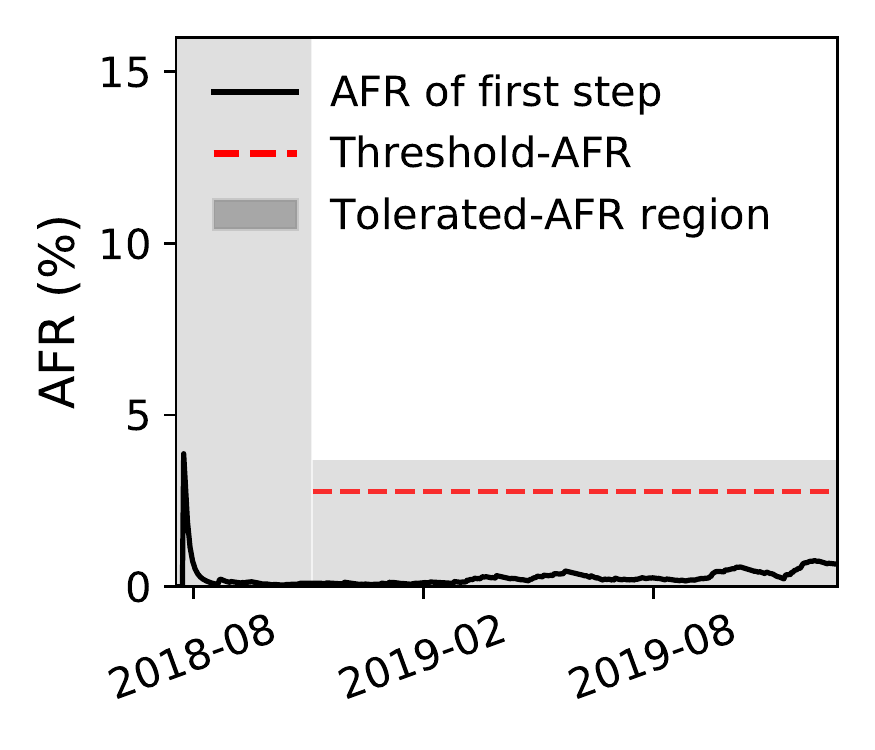}\vspace{-5pt}
        \caption{G-6 (step) \afr curve}
        \label{fig:g_cluster1_afr1}
    \end{subfigure}
    \begin{subfigure}[t]{0.24\textwidth}
        \centering
        \includegraphics[width=\textwidth]{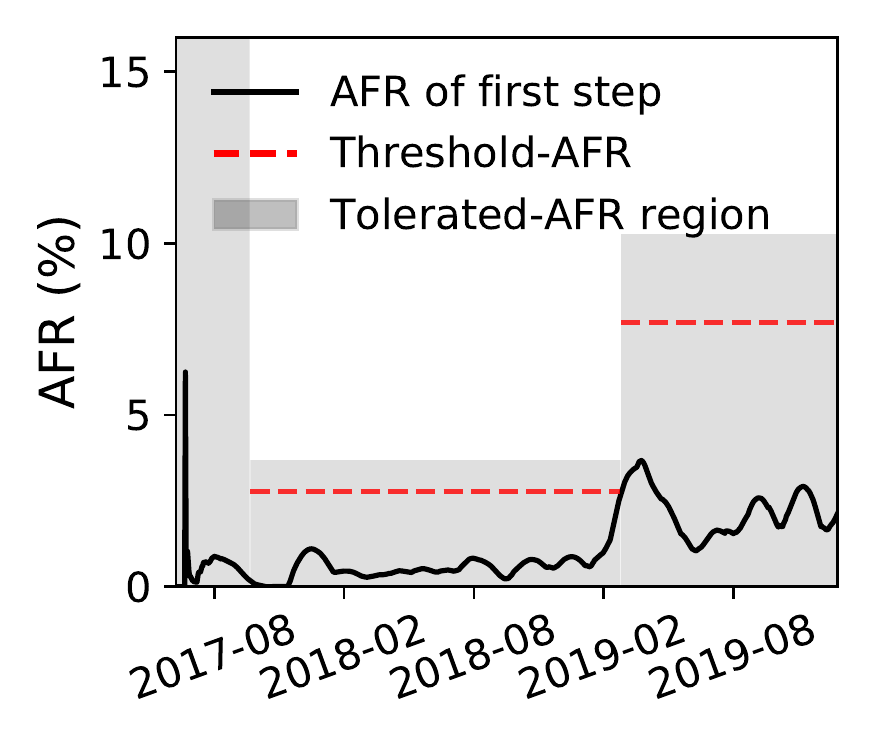}\vspace{-5pt}
        \caption{G-1 (step) \afr curve}
        \label{fig:g_cluster1_afr2}
    \end{subfigure}
    \begin{subfigure}[t]{0.24\textwidth}
        \centering
        \includegraphics[width=\textwidth]{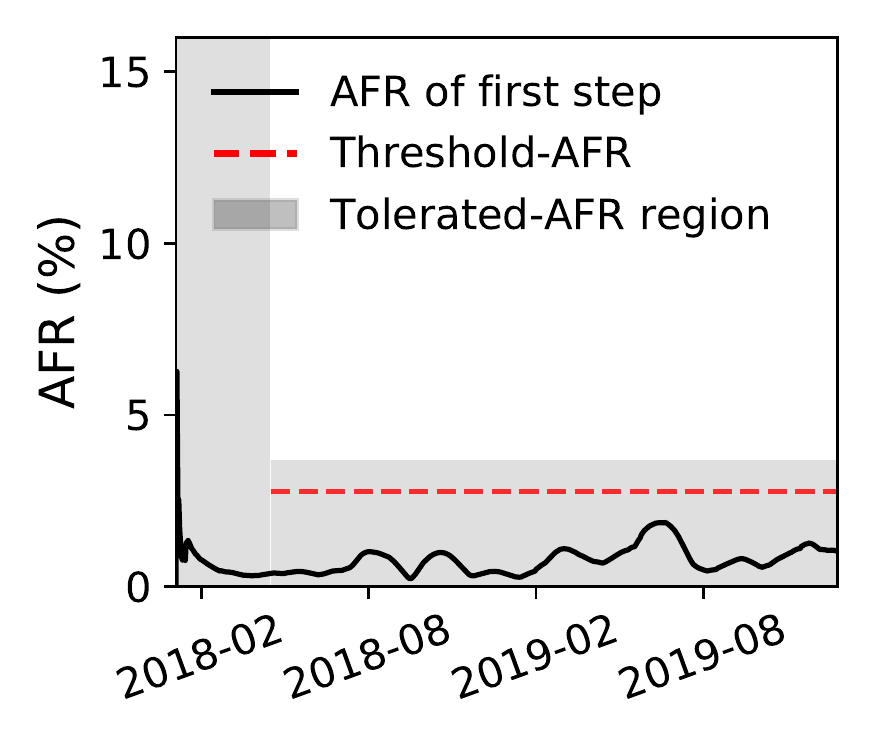}\vspace{-5pt}
        \caption{G-5 (step) \afr curve}
        \label{fig:g_cluster1_afr3}
    \end{subfigure}
    \begin{subfigure}[t]{0.24\textwidth}
        \centering
        \includegraphics[width=\textwidth]{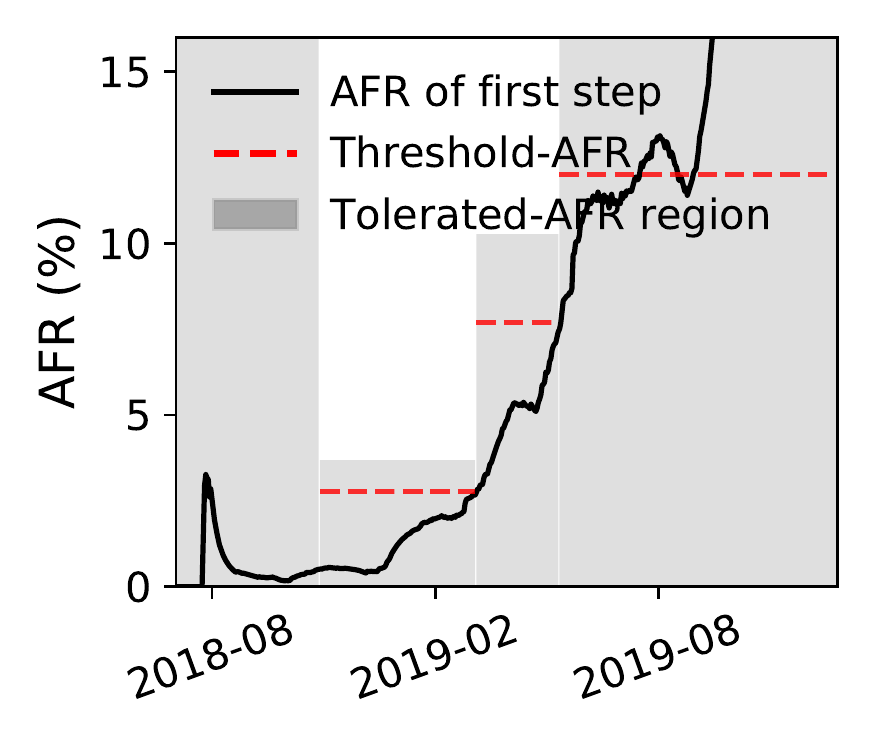}\vspace{-5pt}
        \caption{G-3 (step) \afr curve}
        \label{fig:g_cluster1_afr4}
    \end{subfigure}
    \caption{Detailed IO analysis and space savings achieved by \pacemaker{-enabled} adaptive redundancy on Google Cluster2.}
    \label{fig:main_results_2}
\end{figure*}

\textbf{Google Cluster2.} Fig.~\ref{fig:g_cluster1_after_20} shows the \pacemaker{-generated} IO for redundancy management. Fig.~\ref{fig:g_cluster1_space} shows the corresponding space savings. Finally Figs.~\ref{fig:g_cluster1_afr1}--\ref{fig:g_cluster1_afr4} shows the \afr{s} of the four \adgroup{s} that make up Cluster2. All \adgroup{s} in Google Cluster2 are \stepdeped. Thus, it is not surprising that Fig.~\ref{fig:transition_type} shows that over 98\% of the transitions in Cluster2 were performed by bulk parity recalculation. This is the largest cluster \pacemaker was simulated on. Cluster2's disk population exceeds 450K disks. Even at such large scales, \pacemaker is able to obtain average space savings of almost 17\% and peak space savings of over 25\%. This translates to needing 100K fewer disks, essentially saving millions of dollars.

\begin{figure*}[t]
    \centering
    \begin{subfigure}[t]{0.8\textwidth}
        \includegraphics[width=\textwidth]{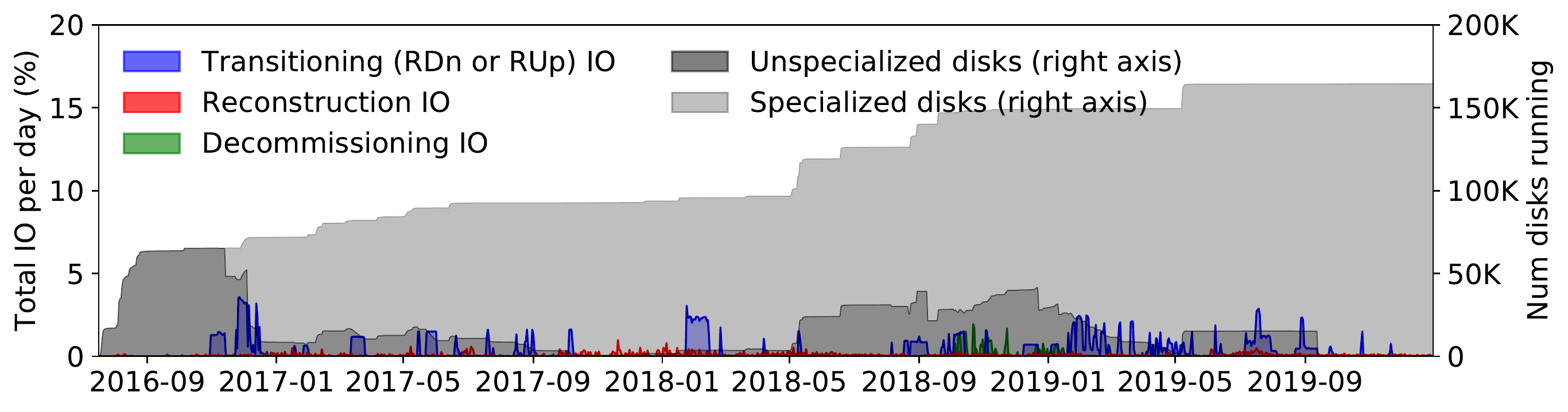}\vspace{-5pt}
        \caption{Google Cluster3 redundancy management IO due to \pacemaker over its 3 year lifetime broken down by IO type.}
        \label{fig:g_cluster2_after_20}
    \end{subfigure}
    \begin{subfigure}[t]{0.76\textwidth}
        \hspace*{-0.85cm}
        \includegraphics[width=\textwidth]{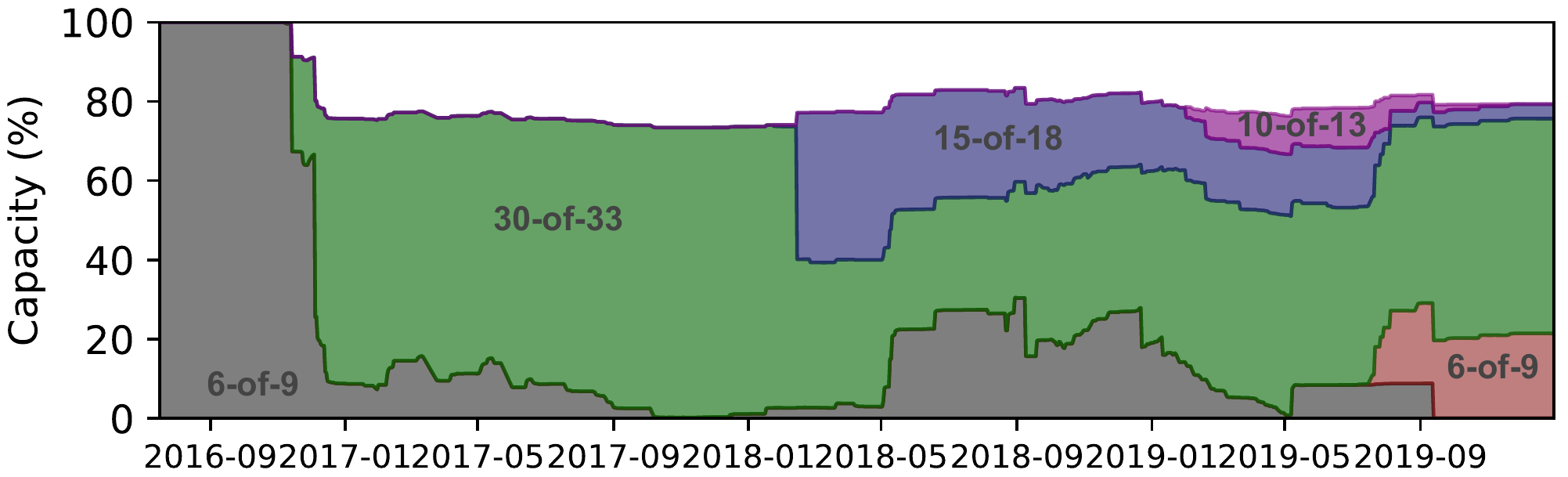}\vspace{-5pt}
        \caption{Google Cluster3 space savings achieved by \pacemaker.}
        \label{fig:g_cluster2_space}
    \end{subfigure}
    \begin{subfigure}[t]{0.24\textwidth}
        \centering
        \includegraphics[width=\textwidth]{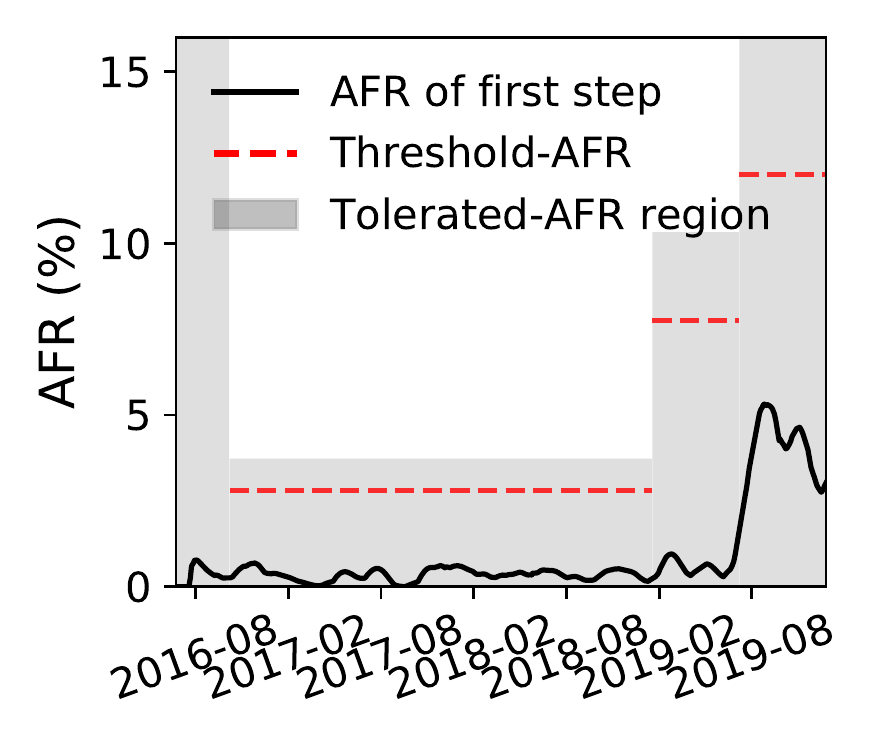}\vspace{-5pt}
        \caption{G-2 (step) \afr curve}
        \label{fig:g_cluster2_afr1}
    \end{subfigure}
    \begin{subfigure}[t]{0.24\textwidth}
        \centering
        \includegraphics[width=\textwidth]{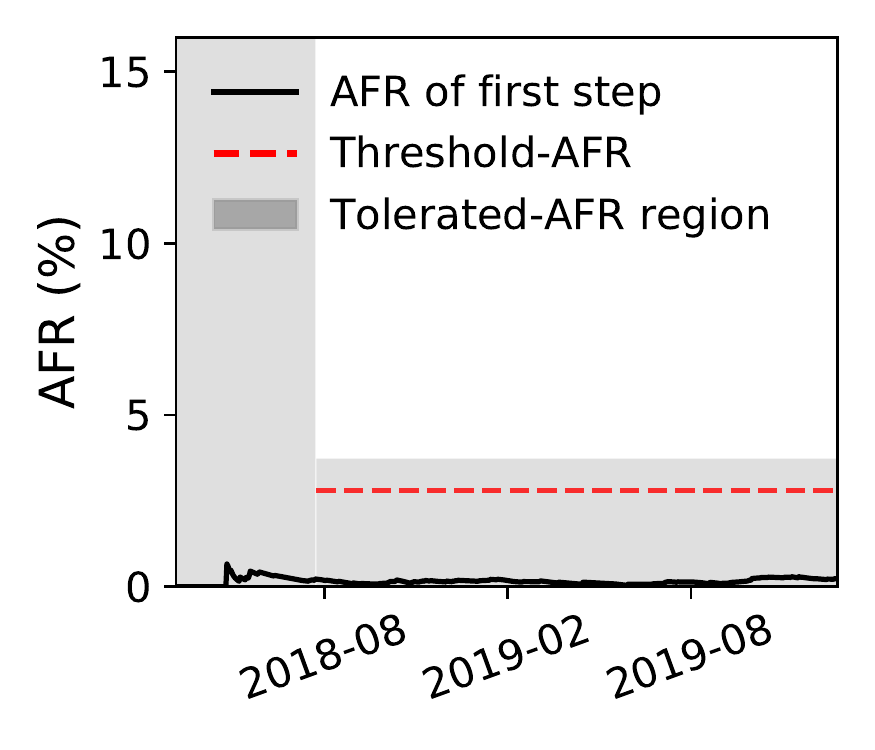}\vspace{-5pt}
        \caption{G-4 (step) \afr curve}
        \label{fig:g_cluster2_afr2}
    \end{subfigure}
    \begin{subfigure}[t]{0.24\textwidth}
        \centering
        \includegraphics[width=\textwidth]{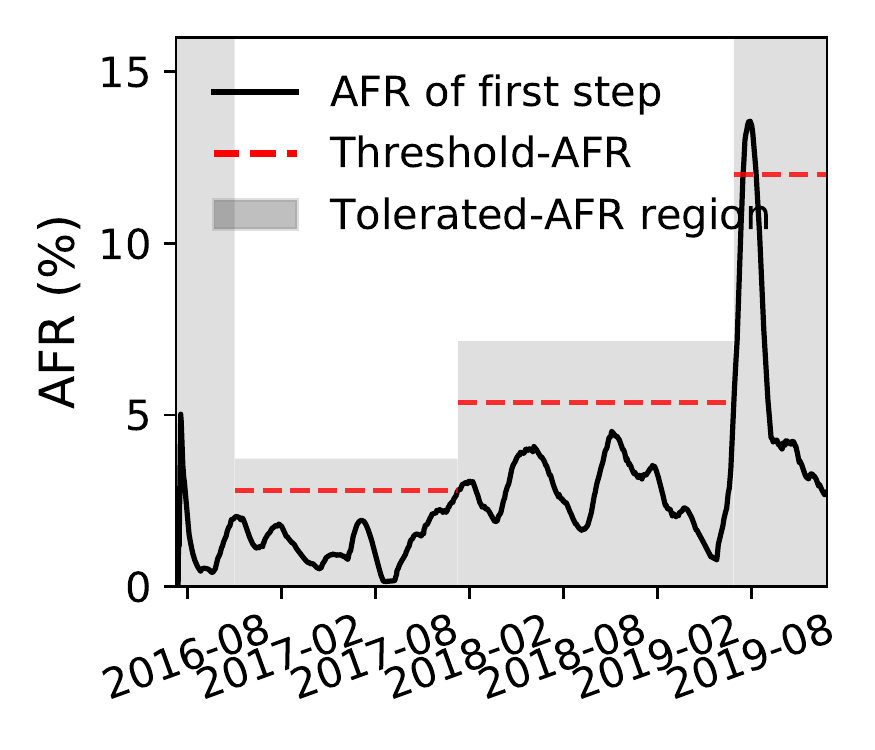}\vspace{-5pt}
        \caption{G-1 (step) \afr curve}
        \label{fig:g_cluster2_afr3}
    \end{subfigure}
    \caption{Detailed IO analysis and space savings achieved by \pacemaker{-enabled} adaptive redundancy on Google Cluster3.}
    \label{fig:main_results_3}
\end{figure*}

\textbf{Google Cluster3.} Google Cluster3 is not as large as Cluster1 or Cluster2. At its peak, Cluster3 has a disk population of approximately 200K disks. But, it achieves the highest average space savings (20\%) compared to all other clusters. Fig.~\ref{fig:g_cluster2_after_20} shows the \pacemaker{-generated} IO, Fig,~\ref{fig:g_cluster2_space} shows the space savings and Figs.~\ref{fig:g_cluster2_afr1}--\ref{fig:g_cluster2_afr3} shows the \afr curves of its three \adgroup{s}. Like Cluster2, Cluster3 is also mostly \stepdeped.

\begin{figure*}[t]
    \centering
    \begin{subfigure}[t]{0.8\textwidth}
        \includegraphics[width=\textwidth]{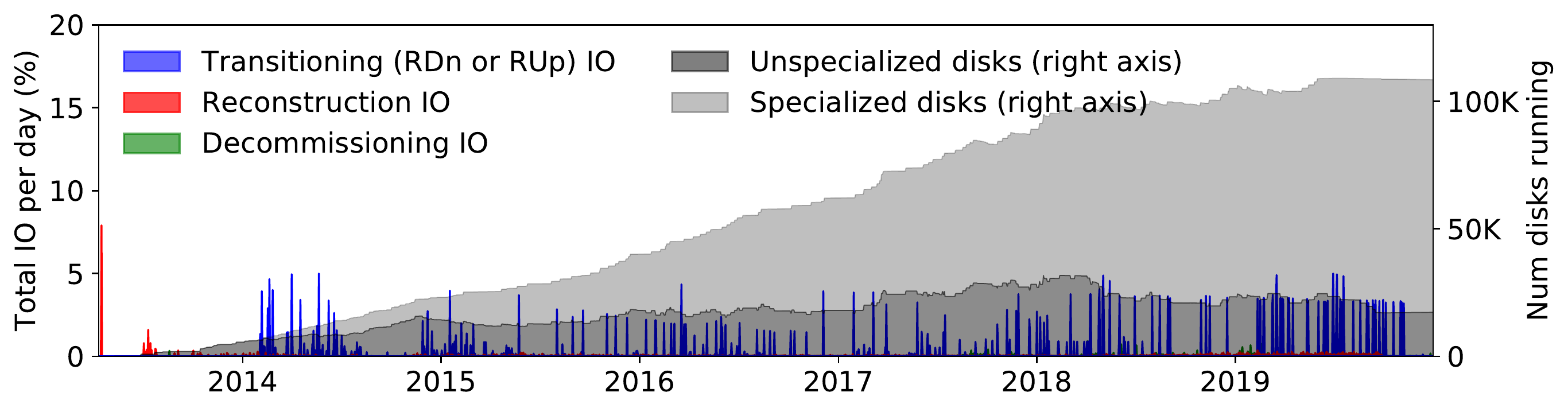}\vspace{-5pt}
        \caption{Backblaze redundancy management IO due to \pacemaker over its 6$+$ year lifetime broken down by IO type.}
        \label{fig:bb_after_20}
    \end{subfigure}
    \begin{subfigure}[t]{0.76\textwidth}
        \hspace*{-0.85cm}
        \includegraphics[width=\textwidth]{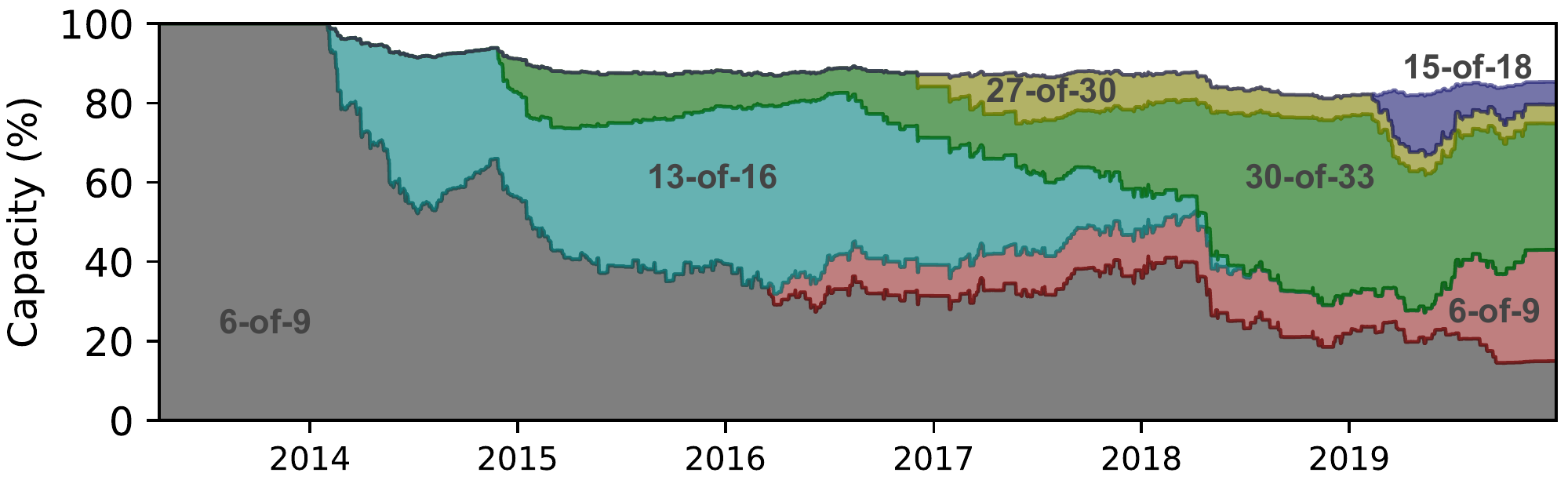}\vspace{-5pt}
        \caption{Backblaze space savings achieved by \pacemaker.}
        \label{fig:bb_space}
    \end{subfigure}
    \begin{subfigure}[t]{0.24\textwidth}
        \centering
        \includegraphics[width=\textwidth]{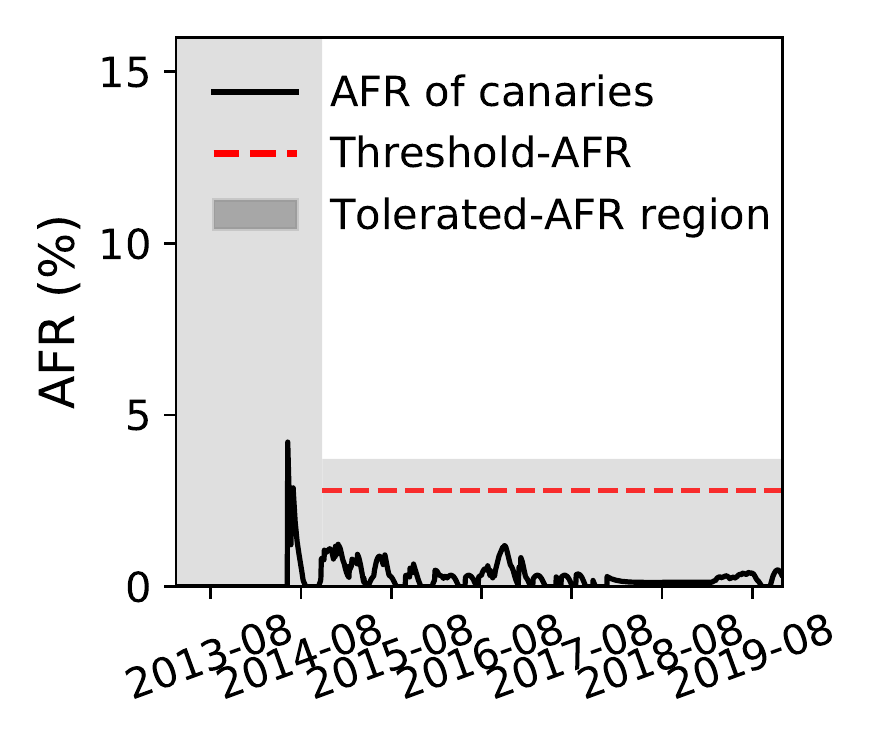}\vspace{-5pt}
        \caption{H-4A (trickle) \afr curve}
        \label{fig:bb_h4a}
    \end{subfigure}
    \begin{subfigure}[t]{0.24\textwidth}
        \centering
        \includegraphics[width=\textwidth]{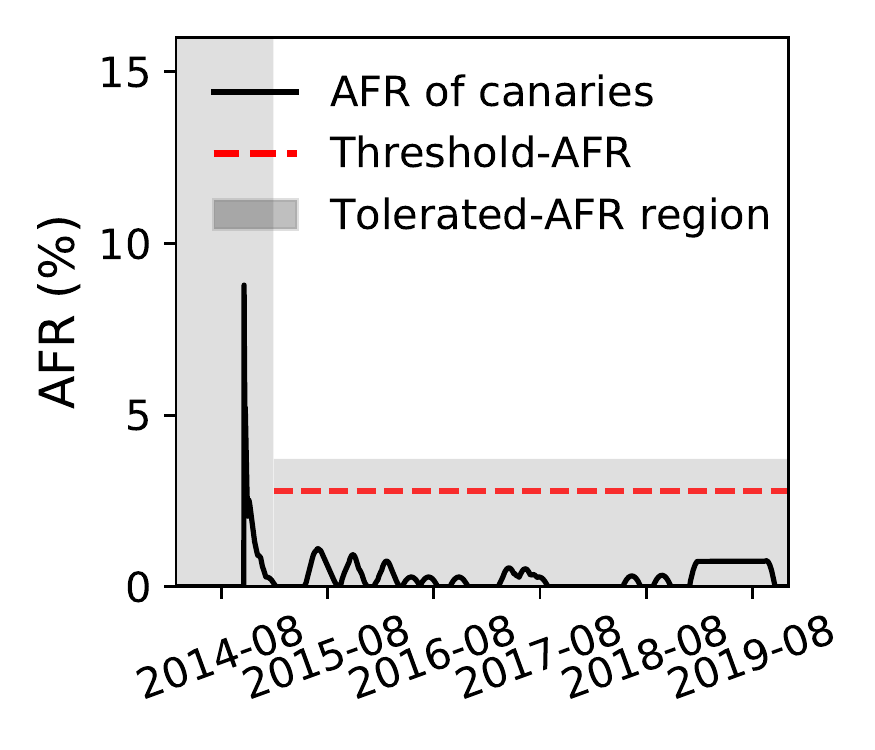}\vspace{-5pt}
        \caption{H-4B (trickle) \afr curve}
        \label{fig:bb_h4b}
    \end{subfigure}
    \begin{subfigure}[t]{0.24\textwidth}
        \centering
        \includegraphics[width=\textwidth]{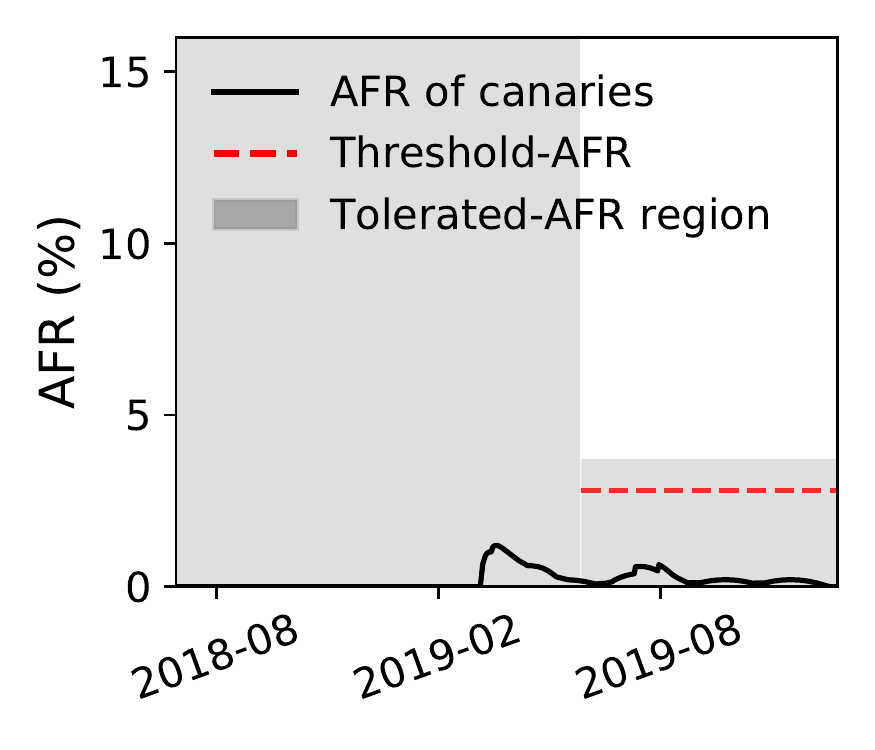}\vspace{-5pt}
        \caption{H-12E (trickle) \afr curve}
        \label{fig:bb_h12e}
    \end{subfigure}
    \begin{subfigure}[t]{0.24\textwidth}
        \centering
        \includegraphics[width=\textwidth]{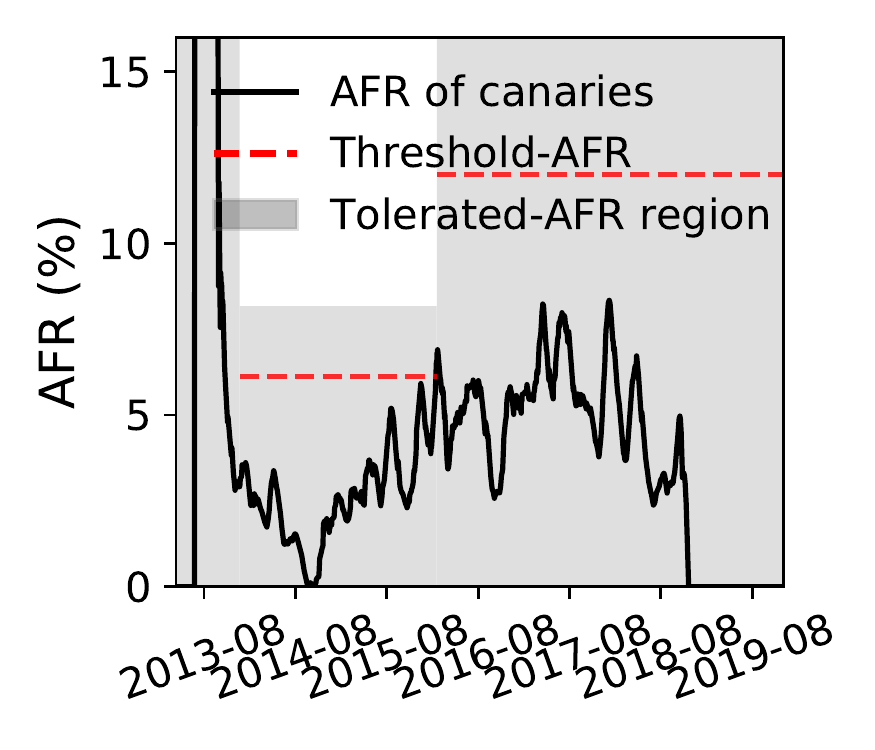}\vspace{-5pt}
        \caption{S-4 (trickle) \afr curve}
        \label{fig:bb_s4}
    \end{subfigure}
    \begin{subfigure}[t]{0.24\textwidth}
        \centering
        \includegraphics[width=\textwidth]{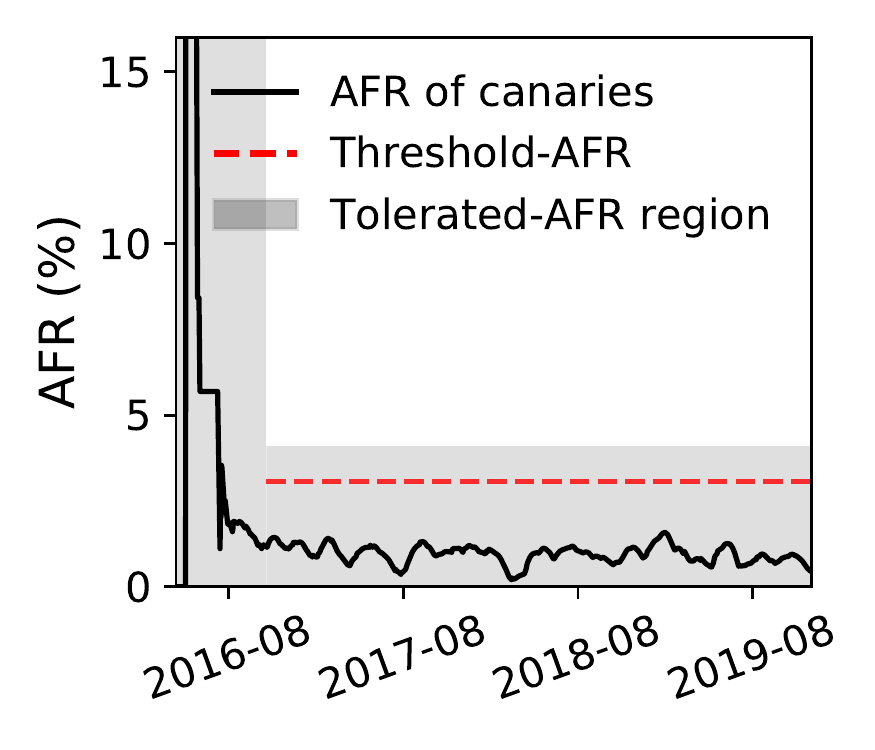}\vspace{-5pt}
        \caption{S-8C (trickle) \afr curve}
        \label{fig:bb_s8c}
    \end{subfigure}
    \begin{subfigure}[t]{0.24\textwidth}
        \centering
        \includegraphics[width=\textwidth]{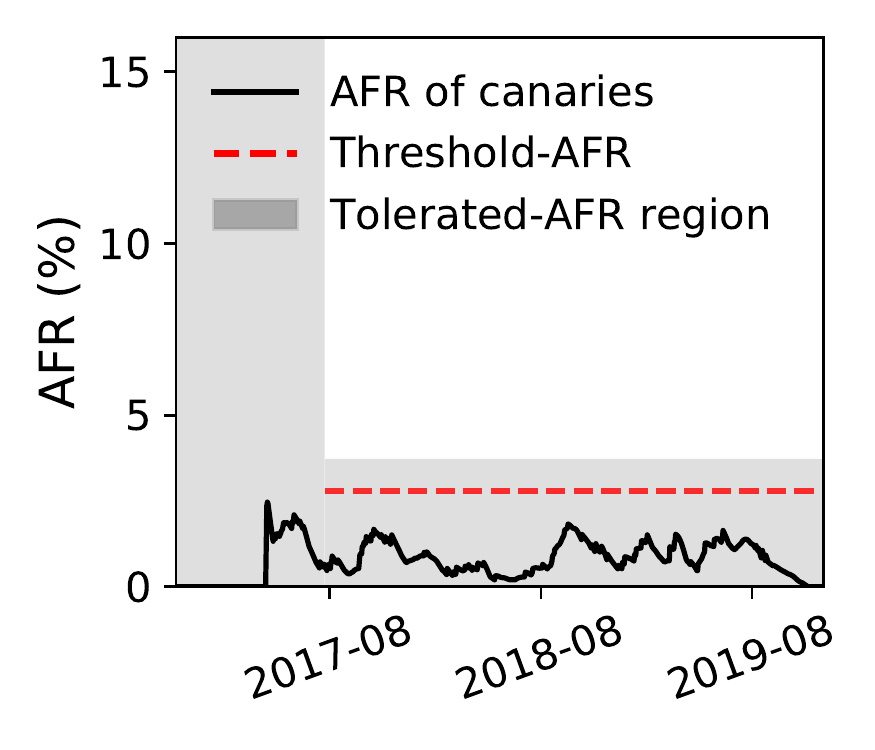}\vspace{-5pt}
        \caption{S-8E (trickle) \afr curve}
        \label{fig:bb_s8e}
    \end{subfigure}
    \begin{subfigure}[t]{0.24\textwidth}
        \centering
        \includegraphics[width=\textwidth]{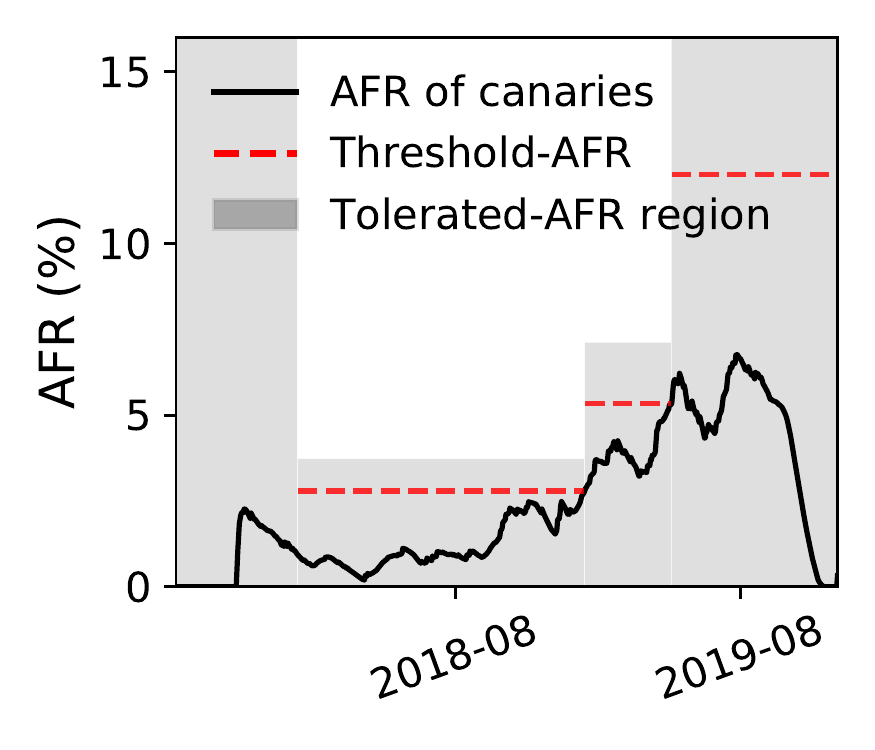}\vspace{-5pt}
        \caption{S-12E (trickle) \afr curve}
        \label{fig:bb_s12e}
    \end{subfigure}
    \caption{Detailed IO analysis and space savings achieved by \pacemaker{-enabled} adaptive redundancy on the Backblaze cluster.}
    \label{fig:main_bb_results}
\end{figure*}

\textbf{Backblaze Cluster.} Backblaze is a completely \trickledeped cluster. Fig.~\ref{fig:bb_after_20} shows the \pacemaker{-generated} IO. Unlike Google clusters, the transition IO of Backblaze does not produce large regions of transition workload. Instead, since \trickledeped disks transition a-few-at-a-time, we see transition work appearing continuously throughout the cluster lifetime of over 6 years. Unsurprisingly, most of the transitions are done by emptying disks (Type 1; refer Fig.~\ref{fig:transition_type}). In terms of sensitivity, the Backblaze cluster is the most insensitive to the \peakio constraint since always requires much lower transition bandwidth per day.